# Dynamic fracture of a bicontinuously nanostructured copolymer: A deep-learning analysis of *big-data-generating experiment*

Hanxun Jin, Tong Jiao, Rodney J. Clifton, and Kyung-Suk Kim*

School of Engineering, Brown University, Providence, RI 02912, United States

* Corresponding author: kyung-suk_kim@brown.edu

**Abstract:**

Here, we report measurements of detailed dynamic cohesive properties (DCPs) beyond the dynamic fracture toughness of a bicontinuously nanostructured copolymer, polyurea, under an extreme loading rate, from deep-learning analyses of a dynamic *big-data-generating experiment*. We first describe a new Dynamic Line-Image Shearing Interferometer (DL-ISI), which uses a streak camera to record optical fringes of displacement-gradient vs time profile along a line on sample's rear surface. This system enables us to detect crack initiation and growth processes in plate-impact experiments. Then, we present a convolutional neural network (CNN) based deep-learning framework, trained by extensive finite-element simulations, that inversely determines the accurate DCPs from the DL-ISI fringe images. For the measurements, plate-impact experiments were performed on a set of samples with a mid-plane crack. A Conditional Generative Adversarial Networks (cGAN) was employed first to reconstruct missing DL-ISI fringes with recorded partial DL-ISI fringes. Then, the CNN and a correlation method were applied to the fully reconstructed fringes to get the dynamic fracture toughness, $12.1\ kJ/m^2$, cohesive strength, $302\ MPa$, and maximum cohesive separation, $80.5\ \mu m$, within $\pm 0.4\%$, $\pm 2.7\%$, and $\pm 2.2\%$ differences, respectively. For the first time, the DCPs of polyurea have been successfully obtained by the DL-ISI with the pre-trained CNN and correlation analyses of cGAN-reconstructed data sets. The dynamic cohesive strength is found to be nearly three times higher than the dynamic-failure-initiation strength. The high dynamic fracture toughness is found to stem from both high dynamic cohesive strength and high ductility of the dynamic cohesive separation. These experimental results fill a gap in the current understanding of nanostructured copolymer's cooperative-failure strength under extreme local loading conditions near the crack tip. This experiment demonstrates the advantages of a *big-data-generating experiment* to extract unprecedented details of nanostructured material's dynamic fracture characteristics with deep-learning algorithms.

# 1 Introduction

Recently, researchers have shown substantial interest in nanoscale dynamic toughening mechanisms of a nanophase-segmented block copolymer, polyurea(Clifton and Jiao, 2015; Grujicic et al., 2015; Jain et al., 2013; Youssef and Gupta, 2012). During the linear polymerization, thermodynamic incompatibility between the polytetramethylene oxide and diisocyanate enables the dual-phase bicontinuous nanostructures consisting of hard gyroid arms dispersed in a continuous soft medium, as observed in Atomic Force Microscopy (AFM) tapping-mode phase images (Das et al., 2007; Grujicic et al., 2014; Jin et al., 2022a; Kim et al., 2020; Wisse et al., 2006). These bicontinuously segregated nanostructures are believed to provide dynamic deformation mechanisms responsible for the ultra-high strength of polyurea under extremely high strain-rate loading. For studying the dynamic failure response of polyurea, release-wave plate-impact experiments (Clifton and Jiao, 2015) were conducted. These experiments revealed a tensile dynamic-failure-initiation stress of approximately 105 MPa for polyurea PU1000 under uniaxial strain loading conditions. However, polyurea did not directly spall within the duration of tensile-wave loading. Instead, micron-size voids were observed on the "spall plane", continuing to transmit stress. Understanding the entire dynamic failure process in the spalling test requires accurate fracture models consisting of void initiation-growth-coalescence processes. Developing such high-fidelity cohesive models requires well-characterized dynamic fracture experiments rather than "spall" tests. Such experiments, involving plane wave loading of samples with a mid-plane crack, have been introduced previously to measure the plane-strain dynamic fracture toughness of AISI 4340 steel, under mode I (Ravichandran and Clifton, 1989). Subsequently, the technique was extended to pressure-shear plate-impact experiments to measure the mode II in-plane shear toughness (Zhang and Clifton, 2003), and the mode III anti-plane shear (Zhang and Clifton, 2007) toughness. A schematic of the mode I dynamic fracture testing experiments using plate-impact loading is illustrated in **Fig. 1(a)**. Although the dynamic mixed-mode cohesive properties of a dissimilar material interface could be measured in general with our methods, we focused on mode I fracture of polyurea, i.e., in this paper, materials A and B in **Fig. 1(a)** are the same.

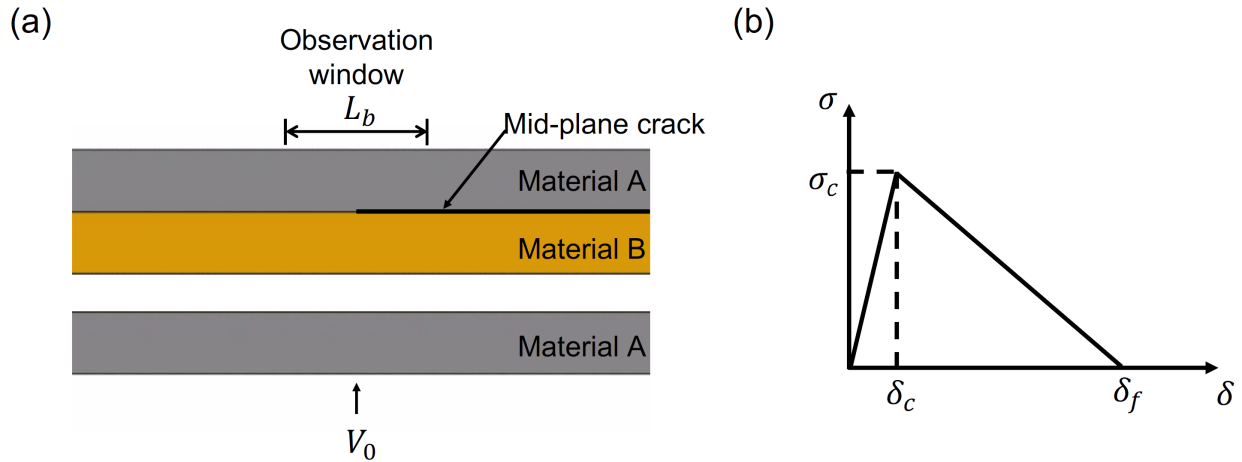

**Figure 1** (a) A schematic of dynamic fracture testing using plate-impact loading: a flyer with initial speed $V_0$ impacts a sample with a mid-plane crack. (b) The schematics of a bilinear cohesive law with three cohesive parameters ($\sigma_c, \delta_c, \delta_f$).

Beyond the fracture-toughness measurements, more information on the strength and ductility in the fracture process zone has been needed to uncover the nano/micro-scale dynamic toughening mechanisms of novel advanced materials. To this end, failure processes are often characterized by cohesive zone models (CZMs) (Barenblatt, 1962; Dugdale, 1960; Tvergaard and Hutchinson, 1996; Xu and Needleman, 1993). In general, the CZM employs a physical traction-separation law to describe the cohesive behavior between similar or dissimilar interfaces. For example, a simple three-parameter bilinear cohesive law is illustrated in **Fig. 1(b)**. Since its development, CZM has been successfully used to describe various fracture and friction problems such as void growth in metals (Tvergaard and Hutchinson, 1996), polymer crazing (Hong et al., 2009; Hong and Kim, 2003), and single and multi-asperity friction (Kim et al., 1998; Li and Kim, 2008, 2009). In contrast to static inverse problems, e.g., (Hong et al., 2009; Hong and Kim, 2003), determining dynamic cohesive laws from experimental data is a more difficult inverse problem.

Recent advances in machine learning, especially deep learning, provide a powerful method to solve inverse problems in engineering and physics (Hsu et al., 2020; Karniadakis et al., 2021; Ni and Gao, 2021; Niu and Srivastava, 2022; Sanchez-Lengeling and Aspuru-Guzik, 2018; Shi et al., 2020; Zhang et al., 2022; Zhang et al., 2020). In principle, the deep-learning approach uses a multi-layer neural network structure that can learn the nonlinear mapping between training dataset and pre-set labels or parameters (LeCun et al., 2015). Then, the pre-trained neural network can effectively predict parameters for the test dataset. With regard to using deep learning to determine the cohesive laws inversely, recent studies (Ferdousi et al., 2021; Su et al., 2021) have used load-displacement curves at the boundary of the specimen to train a deep dense-layer neural-net architecture. However, above-mentioned quasi-static inversion methods cannot be applied to dynamic experiments under extreme conditions like plate-impact experiments where local strain rates exceed $10^6 s^{-1}$. For dynamic-fracture inverse problems, laser interferometric techniques like Normal Displacement Interferometry (NDI) or Transverse Displacement Interferometry (TDI) (Kim et al., 1977) have been used to measure a time trace of displacements at a point to inversely determine the fracture toughness (Ravichandran and Clifton, 1989) based on analytical linear-elastic dynamic crack-tip solutions of Freund (Freund, 1972, 1973). However, such "one-point" measurement cannot accurately determine interfacial properties like the cohesive strength and ductility parameters. A new experimental technique that can generate a sufficient dataset in a single experiment is desired for deep learning to determine the cohesive parameters inversely. We define such a high-throughput experiment as a *big-data-generating experiment*. Such measurement can readily be achieved from our introduction of the Dynamic Line-Image Shearing Interferometer (DL-ISI) with a high-speed streak camera system.

In this paper, we first present a novel Dynamic Line-Image Shearing Interferometer (DL-ISI) coupled with a streak camera system in **Section 2**, which can record displacement-gradient vs time information along a line on the specimen surface. Then, in **Section 3**, we introduce a deep-learning framework that can determine accurate cohesive parameters in dynamic *big-data-generating experiments*. Detailed plate-impact experiments on polyurea with a mid-plane crack are discussed in **Section 4**. The experimental fringe image collected by the streak camera is cleaned and inpainted by a state-of-the-art Conditional Generative Adversarial Network (cGAN). The dynamic cohesive parameters and fracture toughness are then readily determined from the correlation method as well as the deep learning framework. The underlying molecular-level dynamic toughening mechanisms of polyurea will be presented in a sequel (Jin et al., 2022b).

## 2 Dynamic Line-Image Shearing Interferometer (DL-ISI)

Due to the resolution limitation of the streak camera system, recording displacement profiles directly using an interferometer of the Michelson type is challenging. The number of fringes obtained from the Michelson interferometer is $n = 2d/\lambda$, where $d$ is the rear surface displacement, and $\lambda$ is the wavelength of the laser. From a FEM analysis, $d$ is estimated to be of the order of 1mm when the crack opens, and the fringe number $n$ reaches approximately 4000, which is beyond the resolution of the streak camera system with $1024 \times 1024$ pixels. Therefore, a new interferometer that can measure the rear surface line profile with reasonable fringe density is needed. For this purpose, we introduced the Dynamic Line-Image Shear Interferometer (DL-ISI), where the fringe information represents the rear surface's displacement gradients.

The DL-ISI optical design is shown schematically as black lines in **Fig. 2**. In the optical circuit, a laser beam with wavelength $\lambda = 532 nm$ is controlled to synchronously switch the beam on by an acousto-optic modulator (AOM) (Brimrose TEM-110-5-532) located at the focal plane of the spherical lens SL1, for the exposure duration of a high dynamic range streak camera (Hamamatsu C7700). Then the laser beam is collimated and passing through a slit with a size of $10mm \times 2mm$. The slit of collimated beams is illuminated on the rear surface of the specimen with a mirror-quality finish. The orientation of the illuminated slit is perpendicular to the crack-front line. Then, the reflected beam is projected onto the image plane of the streak camera for one-to-one magnification imaging through the two spherical lenses SL2 and SL3 with the same focal length $f_{len}$. At the focal point of SL2, a microscope coverslip glass with $150 \mu m$ thickness is used to create image shearing interfered fringes. The schematics of these two interfered beams reflected from the front surface and rear surface of the glass is shown in **Fig. 3**. The electric field of the beam reflected from the front surface of the glass and arriving at the image plane can be written as,

$$\boldsymbol{E}^{(1)}(x, y, z, t) = \boldsymbol{E}_0^{(1)} exp\left\{i\left[kz + \frac{k}{2R(z)}(x^2 + y^2) - 2ku(x,t) - \omega t\right]\right\}, \qquad (1)$$

where $(x, y)$ is the in-plane coordinate and $z$ the light-propagation coordinate of the image plane, $R(z)$ is the radius of curvature of the wave front at $z$, and $R(z) = f_{len}$ at the image plane. Here, $k$ and $\omega$ are the wavenumber and the angular frequency of the light, $t$ is the time, and $u(x,t)$ is the out-of-plane displacement of the sample. Since the slit height is much smaller than the width (less than 1%), we consider the streak camera to record line displacement information only. The electric field amplitude of the beam reflected from the rear surface of the glass is,

$$E^{(2)}(x + \Delta x, y, z', t)$$

$$= E_0^{(2)} \exp\left\{i\left[kz' + \frac{k}{2R(z')}((x + \Delta x)^2 + y^2) - 2ku(x + \Delta x, t) - \omega t\right]\right\}, \quad (2)$$

where $z' = z + D + \beta z$ with $D$ the optical path difference of these two beams in the glass and $\beta$ is the angle between two reflected beams due to a small tilt angle $\alpha$ in the glass. $\Delta x$ is the shearing distance in $x$ direction. Their relation can be derived as,

$$D = 2h_g(n_g^2 - \sin^2\theta_{in})^{1/2}, \quad (3a)$$

$$\Delta x = 2h_g \frac{\sin\theta_{in}\cos\theta_{in}}{\sqrt{n_g^2 - \sin^2\theta_{in}}}, \quad (3b)$$

$$\beta = \frac{2\alpha(n_g^2 - \sin^2\theta_{in})^{\frac{1}{2}}}{\cos\theta_{in}}, \quad (3c)$$

where $h_g$ is the glass thickness, $\theta_{in}$ is the beam incident angle, and $n_g$ is the refractive index of the glass. The intensity of these two interfered beams is,

$$I(x, y, z, t) = I_1 + I_2 + 2\sqrt{I_1 I_2}\cos(\phi), \quad (4)$$

where the phase difference $\phi$ is,

$$\phi = k\left\{z + \frac{(x^2 + y^2)}{2R(z)} - 2u(x, t) - z' - \frac{[(x + \Delta x)^2 + y^2]}{2R(z')} + 2u(x + \Delta x, t)\right\}. \quad (5)$$

Since the distance from the glass to the image plane, $z$ is 3 orders of magnitude larger than the optical path difference $D$, and the glass tilt angle $\alpha$ is measured as $0.018 mrad$ which is considered as negligible, we consider $R(z) \cong R(z + D + \beta z)$. Assuming $I_1 \cong I_2 = I_0$, the intensity of these two interfered beams at the image plane can be derived as,

$$I(x, t) = 4I_0 \cos^2\left(\frac{\phi}{2}\right) = 4I_0 \cos^2\left\{\frac{\pi}{\lambda}\left[D + \frac{\Delta x}{f_{len}}x - 2\frac{\partial u(x, t)}{\partial x}\Delta x\right]\right\}. \quad (6)$$

As we can see from **Eq. 6**, the DL-ISI fringe variation depends on the displacement gradient along the image shearing direction $\frac{\partial u(x,t)}{\partial x}$. The DL-ISI fringe sensitivity to displacement gradient, $S$, i.e., displacement gradient per fringe is,

$$S = \frac{\partial u(x,t)}{\partial x}/fringe = \frac{\lambda}{2\Delta x}, \tag{7}$$

The second term in **Eq. 6** represents an initial fringe density which caused by the shearing of the spherical wavefront at the glass plate. The number of initial fringes, $N_{inif}$ at the image plane of the streak camera is,

$$N_{inif} = \frac{L_b \Delta x}{f_{len}\lambda}, \tag{8}$$

where $L_b$ is the laser-illumination window length as shown in **Fig. 1(a)**. The initial fringe density is defined as,

$$\rho_{fringe} = \frac{\Delta x}{f_{len}\lambda} \tag{9}$$

The DL-ISI fringe sensitivity and initial fringe density can be controlled by the image shearing distance $\Delta x$, which depends on the glass thickness $h_g$ and the incident angle of the laser beam $\theta_{in}$, as shown in **Eq. 3b**. Conventionally, when a dense fringe is obtained in an experiment, fringe phase unwrapping techniques are used to extract the field information. Therefore, the displacement gradients profile can be obtained by unwrapping the fringe image. However, the accuracy of fringe unwrapping is subjected to the fringe smoothness and the unwrapping algorithms. If we could adjust the experimental parameters like glass thickness $h_g$ to obtain fringe density according to the optimized deep-learning performance, we could maximize the prediction performance and eliminate the unnecessary computations on unwrapping data processing. Therefore, in this paper, we emphasize using the fringe image itself for deep-learning analysis rather than using the unwrapped dataset from the fringe image. We adjusted the experimental parameters to obtain the optimum fringe density that provides the most accurate cohesive-parameter determination.

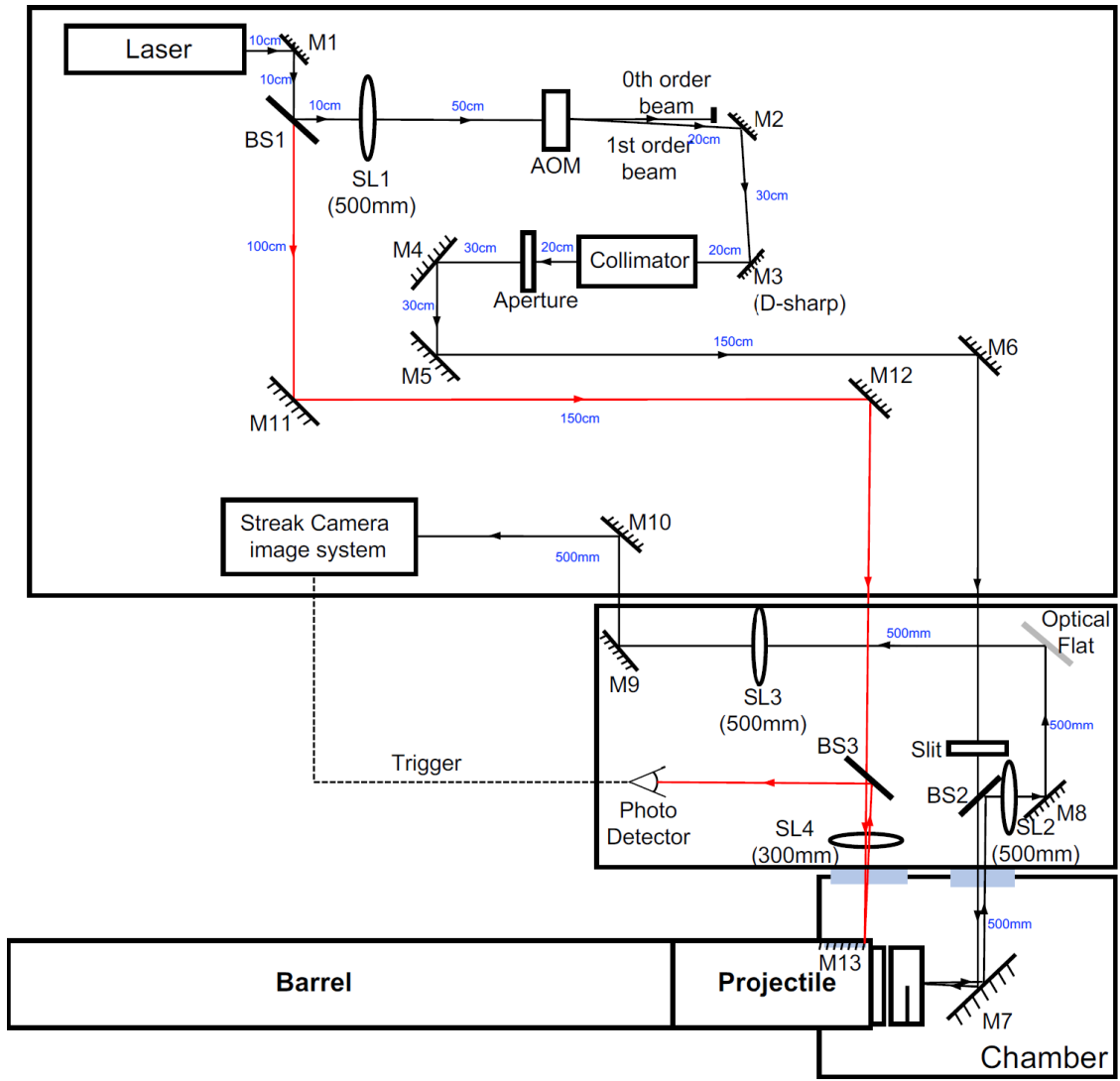

**Figure 2** Optical schematics of dynamic line-image shearing interferometer (DL-ISI) in the plate impact experiment. The black solid line represents the DL-ISI optical circuits, the red solid line represents the optical triggering circuits for the streak camera system. SL represents the spherical Lens; M represents the mirror.

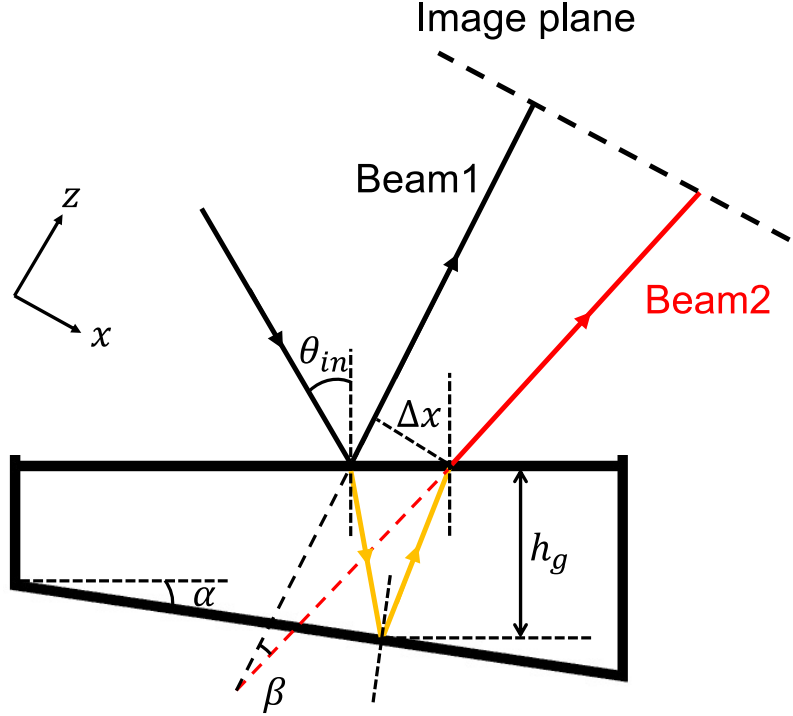

**Figure 3** Schematics of two beams reflected from the front surface and rear surface of a glass plate with thickness $h_g$ and a small tilt angle $\alpha$.

## 3  A deep-learning framework to inversely determine cohesive parameters from *dynamic big-data-generating experiments*

In this section, we present a deep-learning framework that can determine inversely the accurate cohesive parameters in dynamic *big-data-generating experiments*. The general deep learning framework is illustrated in **Fig. 4**. The key components of the framework include: (I) a new DL-ISI experimental technique to collect sufficient experimental data in a single experiment; (II) a computational simulation method to obtain a training dataset within a reasonable time span; (III) a deep-learning algorithm that has the capability to train the network and validate the prediction.

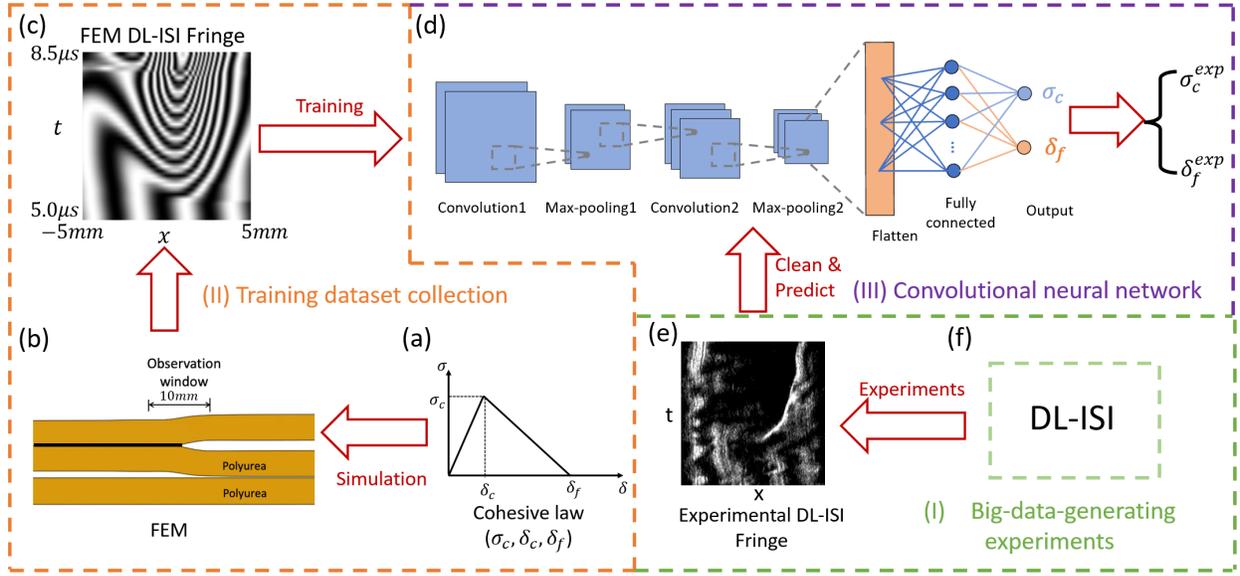

**Figure 4** A deep learning framework to inversely determine cohesive parameters from dynamic *big-data-generating experiments*. (a) A linear traction-separation cohesive laws with three parameters ($\sigma_c, \delta_c,$ and $\delta_f$). (b) Finite element simulations with the interfacial cohesive laws. (c) The Dynamic Line-Image Shearing interferometry (DL-ISI) fringes are generated from FEM. The horizontal axis represents spatial information, while the vertical axis represents temporal information. (d) A convolutional neural network (CNN) is composed of two convolutional layers, two max-pooling layers, one flattened layer, and one hidden layer for two outputs ($\sigma_c, \delta_f$). (e) The experimental DL-ISI fringes were generated from the streak camera system. (f) Dynamic plate impact experiments with DL-ISI setups.

### 3.1 Dataset preparation from FEM simulations

Training a deep-learning neural network usually requires an adequate training dataset. In our case, the input variables would be a DL-ISI fringe image, and the output variables would be three cohesive parameters ($\sigma_c, \delta_c, \delta_f$) of a bilinear interfacial cohesive law (**Fig. 4(a)**). Since the dynamic crack initiation and growth depends primarily on the fracture toughness of the interface, $G = \frac{1}{2}\sigma_c \delta_f$, we consider the two cohesive parameters ($\sigma_c, \delta_f$) as the output variables and set $\delta_c = 0.2\delta_f$ for simplicity. The relevant parameter space of the output variables is selected as $\sigma_c \in$ $[150, 350]\ MPa$ and $\delta_f \in [50, 150]\ \mu m$, which yields the interfacial fracture toughness $G \in [3.75,\ 26.3]\ kJ/m^2$. Initially, we selected $\sigma_c \in [100, 300]\ MPa$, which covers dynamic-failure-initiation strength of homogeneous microcrack nucleation in the bulk material. We found that the computational fringes generated near the lower bound of $\sigma_c$ (i.e., $100\ MPa$) had very low correlation with the experimental fringes. As we searched for the cohesive parameters for maximum correlation, we narrowed the search range to $\sigma_c \in [150, 350]\ MPa$ for higher prediction accuracy. The parameter-space range difference showed a deep-learning prediction difference of about 2%. The range of the separation $\delta_f$ was determined based on the craze dimension estimated from experimental postmortem fracture surfaces. Therefore, we think this parameter range is representative of the cohesive laws for polyurea. The parameter space is uniformly discretized into $m \times m$ intervals. Therefore, the total number of training samples is $M = (m+1)^2$. We chose $m = 9, 13, 19, 24, 30$, which yields $M = 100, 196, 400, 625, 961$, to explore the effect of the training dataset on the neural network performance.

Next, we carried out FEM simulations to obtain the training dataset, i.e., the computational fringes with the labeled cohesive parameters. The FEM simulations on plate impact experiments were performed with the ABAQUS/Explicit 2017 package. As shown in **Fig. 4(b)**, a flyer made of polyurea impacts a polyurea sample with a mid-plane crack. The contact interaction between the flyer and the sample was defined as "hard contact" and frictionless. The simulations were conducted under 2D plane-strain conditions. To reduce the effects of boundary waves on the surface motion, the simulation domain was chosen as the same size as the experimental samples, i.e., the length is 50mm. The impact speed of the flyer is defined as an initial condition of 204m/s. The left half portion of the interface is considered bonded with a cohesive law. Each cohesive law with two cohesive parameters $(\sigma_c, \delta_f)$ was implemented through a VUINTER subroutine of the ABAQUS in the simulation. The strain energy of polyurea can be expressed as, $\overline{W}(\overline{I}_1, J) = C_{10}(\overline{I}_1 - 3) + f(J)$, where $f(J) = -A(J^{-N-1} - J^{-M-1})$ (Clifton and Jiao, 2015), where $\overline{I}_1$ is the invariant of the left Cauchy-Green tensor, and $J$ is the volume ratio between the current and reference volumes. Here, $C_{10} = 150\ MPa$, $A = 858\ MPa$, $N = 6$, $M = 3$, and the density of polyurea is 1070 $kg/m^3$. The detailed constitutive law and parameter selection for polyurea is given in (Clifton and Jiao, 2015). This physics-based instantaneous finite deformation constitutive law with quasi-linear viscoelasticity represents the mechanical behavior well for high-pressure impacts, such as the quasi-isentrope behavior up to 18 GPa. However, the model without quasi-linear viscoelasticity is more accurate for relatively low-pressure (less than 300 MPa) impact, as discussed in (Clifton and Jiao, 2015). We employed the model without quasi-linear viscoelasticity for FEM simulations of our low-pressure impact experiments. Polyurea was modeled through a user subroutine VUMAT in the ABAQUS. The simulation domain is filled with approximately 50,000 gradient-mesh elements, where the smallest element size around the crack tip is 50 $\mu m$. The element type is CPE4R, which is plane strain 4-node bilinear hybrid element with reduced integration. All solutions reported here are the final converged solutions checked by mesh sensitivity tests. The average wall time of each simulation was 10s when running on a CPU with 24 cores. The total training dataset collection for 961 simulations took ~3 hours. After each simulation, the displacement profile along a line within the 10mm observation window at the middle of the target's rear surface was extracted to ensure the area of interest was consistent with the experiments. Then, displacement gradients along the line were calculated to plot corresponding computational DL-ISI fringe images based on the theoretical derivations in **Section 2**. Each simulated DL-ISI fringe image was then cropped into a time window of interest, i.e., the initial crack-tip opening/running/arrest period and rescaled into 64 × 64 pixels for training. An example of a cropped computational fringe image is shown in **Fig. 4(c)**.

To explore the fringe sensitivity to the detailed cohesive laws, the DL-ISI fringes from 3 different cohesive laws: bilinear, trapezoid, and exponential laws, with the same fracture toughness, were plotted in **Fig. A.1** in **Appendix A**. As we can see, the fringe images are similar, and the image correlations among them are larger than 0.98. Therefore, our DL-ISI fringes are mainly governed by fracture energy, with less sensitivity to the detailed cohesive laws. However, the wall time for each FEM simulation with the exponential cohesive law is almost twice longer than the bilinear cohesive law. Hence, in this paper, we use the bilinear traction-separation cohesive law for computational efficiency.

## 3.2 Convolutional neural network (CNN)

The recent development of convolutional neural network (CNN) has shown a compelling capability to classify images into thousands of labels (Krizhevsky et al., 2012). These developments are having enormous impact on modern technology (e.g., face recognition (Parkhi et al., 2015) and self-driving cars (Bojarski et al., 2016)). Here, we adopted a CNN-based neural network to train our computational fringe images with labeled cohesive parameters. The fringe images were first fed into a convolution layer with the rectified linear activation function (ReLU), followed by a conventional 2 × 2 max-pooling layer. Then, the output was flattened and fed into a fully connected layer for two outputs $\sigma_c$, and $\delta_f$. To validate and test the CNN network, the total dataset was randomly divided into 80% training dataset, 10% validation dataset, and 10% test dataset. The accuracy metrics in estimating cohesive parameters were represented by the Mean Absolute Percentage Error (MAPE) as,

$$MAPE = \frac{1}{N}\sum_{i=1}^{N}|\frac{y_i^{true} - y_i^{pre}}{y_i^{true}}| \times 100\%. \tag{10}$$

Here $N$ is the number of the dataset elements, and $y_i^{true}, y_i^{pre}$ are ground-true values and predicted values for the cohesive parameters. The CNN optimized the Mean Square Error (MSE) loss by using the Adam optimizer (Kingma and Ba, 2014) with the learning rate of 0.0001 for 2000 epochs on the open-source platform TensorFlow V2.5 (Abadi et al., 2016). The total training time for the dataset with 961 fringe images on an 8-core CPU is approximately 5 mins.

### 3.3 Train and test

To validate the training process as well as prevent unexpected training issues such as overfitting, the MAPE of training and validation datasets for $\sigma_c$ and $\delta_f$ as a function of the training epoch is plotted in **Fig. 5(a)&(b)**, respectively. As we can see in **Fig. 5**, the MAPEs of both training and validation datasets drop rapidly to 2% when the training epoch reaches ~250, which validates the effectiveness of the current architecture. The MAPEs of both $\sigma_c$ and $\delta_f$ gradually decrease to targeted 1% as the training epoch increases. It is worth noting that the MAPE for validation datasets is slightly higher (~0.5%) than the train datasets. This discrepancy exists for all dataset sizes and CNN architectures we considered. It may be due to the limitation of fringe sensitivity to cohesive parameters. Therefore, decreasing the train dataset mesh to increase the train dataset size does not help. Nevertheless, the overall performance of the current CNN architecture with MAPE less than 1% has already surpassed many exiting inverse algorithms for cohesive parameter determinations for nonlinear materials.

To observe the prediction performance on an individual case, the predicted and ground-truth cohesive parameters are plotted in **Fig. 6(a)&(b)** for the training and testing datasets, respectively. The coefficients of determination ($R^2$) between predicted and true values were also calculated to illustrate the training and prediction performance. With the 2-block CNN architecture, the $R^2$ values on the training dataset are greater than 0.999 for both $\sigma_c$ and $\delta_f$. The $R^2$ values on the test dataset for both parameters are greater than 0.997, which illustrates that the current deep learning architecture can inversely identify the cohesive parameters from the given fringe images, and there is no overfitting on the training datasets.

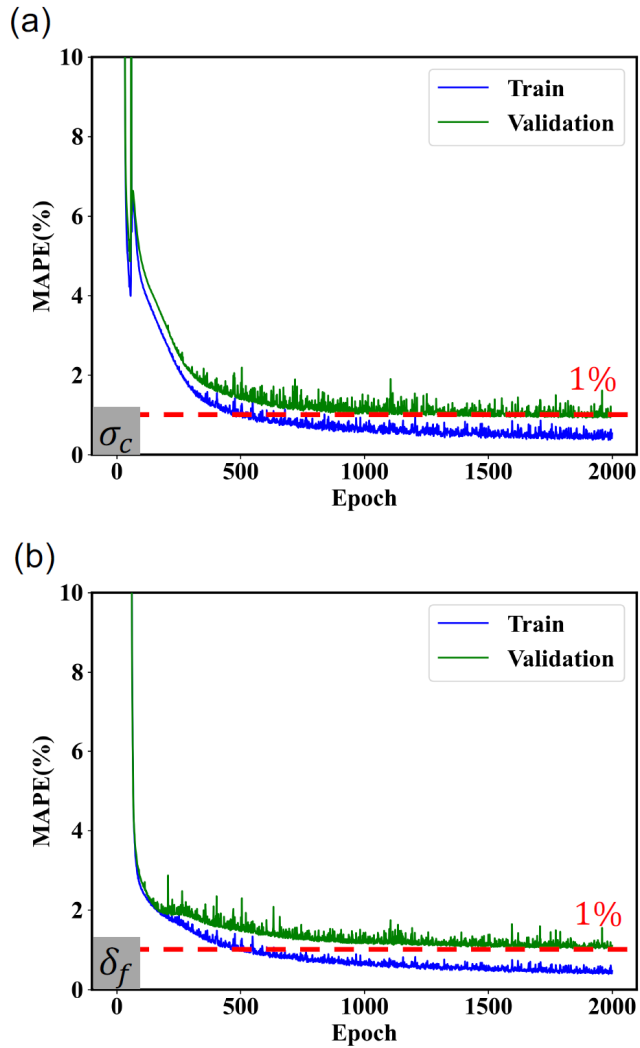

**Figure 5** The Mean Absolute Percentage Error (MAPE) of training and validation datasets as a function of train epoch for (a) $\sigma_c$, and (b) $\delta_f$ with total number of 961 FEM datasets.

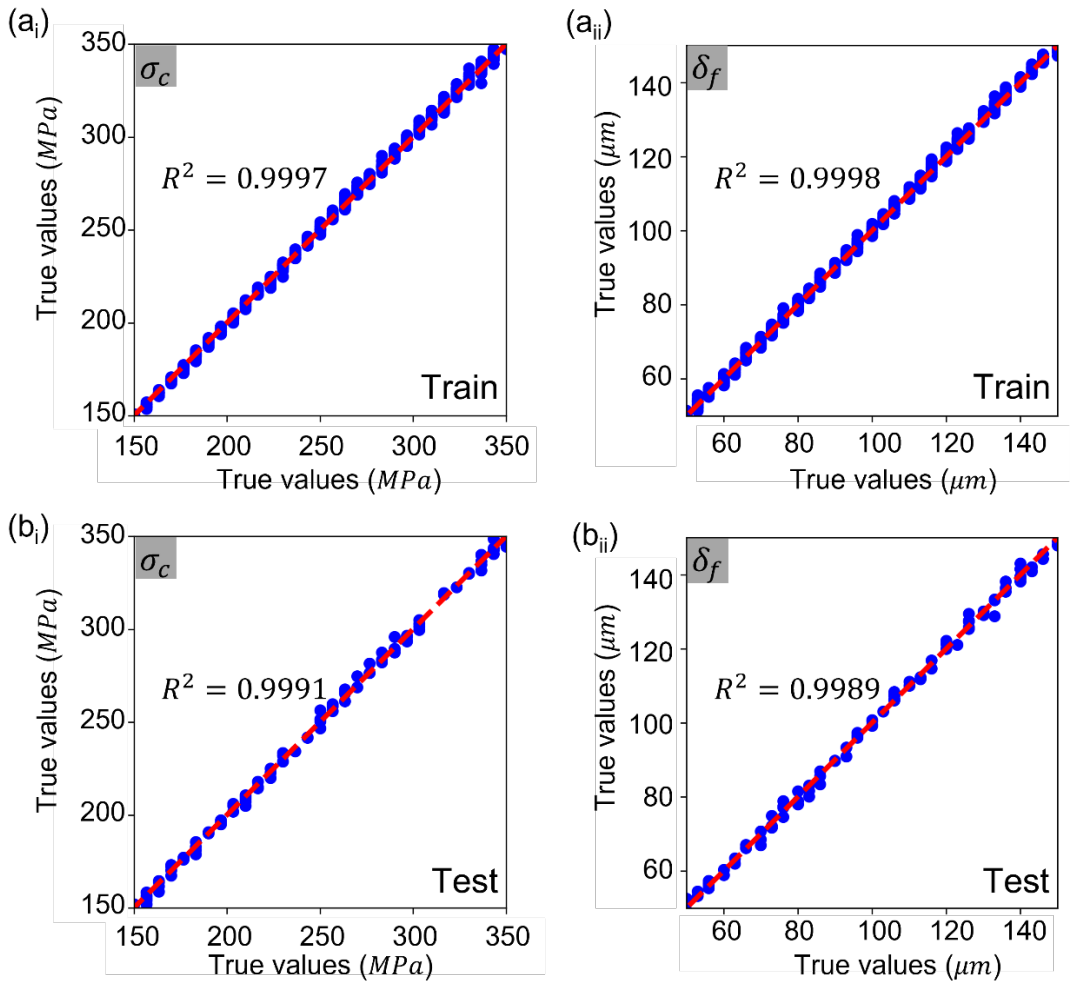

**Figure 6** Prediction preformation of cohesive parameters on training datasets for (a_i) $\sigma_c$, and (a_ii) $\delta_f$, and testing datasets for (b_i) $\sigma_c$, and (b_ii) $\delta_f$.

# 4 Experiments

## 4.1 Design of polyurea specimen with mid-plane crack

Making a polyurea sample with a sharp and straight mid-plane crack is crucial in the plate impact experiment. We made the sample through a new injection method. First, a polyurea solution was prepared from the mixture of an oligomeric diamine, Versalink P-1000 (Air Product) with a diisocyanate, Isonate 143L (Dow Chemicals) in a 4:1 ratio. The P-1000 and Isonate 143L mixture was stirred by a glass stirring rod for 1 min at ambient conditions and degassed for 6 mins in a vacuum chamber. Then the degassed mixture was poured into a syringe. As shown in **Fig. 7(a)**, the syringe symmetrically injected the mixture into two sides of an acrylic mold which was coated with polyurea release spray. A Teflon thin film with a thickness of $50 \mu m$ was firmly clamped in the bottom half of the acrylic frames and the supportive plates. As the injection proceeded, the mixture automatically lifted the supportive plates while keeping the Teflon thin-film straight. The polyurea specimen was cured and stored in a desiccator at room temperature for at least 72 hours before demolding. The demolded specimen was cut into $50 \ mm \times 50 \ mm$ squares with mid-plane crack aligned in the middle of the sample, as shown in **Fig. 7(b)**. In this way, a polyurea specimen was prepared to have a sharp mid-plane crack and a glossy surface for an aluminum coating. The polyurea flyer was directly molded to a disk shape with a $50 \ mm$ diameter. The flyer and the specimen were molded from the same batch of mixture solution to ensure their material properties were the same. Both flyer and specimen were coated with 200 nm aluminum coating for conductivity for the tilt triggering and reflectivity for the DL-ISI measurement. After coating, both flyer and specimen were checked by Newton's ring method to ensure the surface flatness was within 3.0 $\mu m$ over 2 inches. Parallelity of the sample was determined by the acrylic spacer, which had 0.33 mrad parallelity. The Teflon thin film was removed for the first two shots, but it was found the compressive wave will partially self-heal the mid-plane crack. Therefore, the Teflon thin film was kept in the third shot.

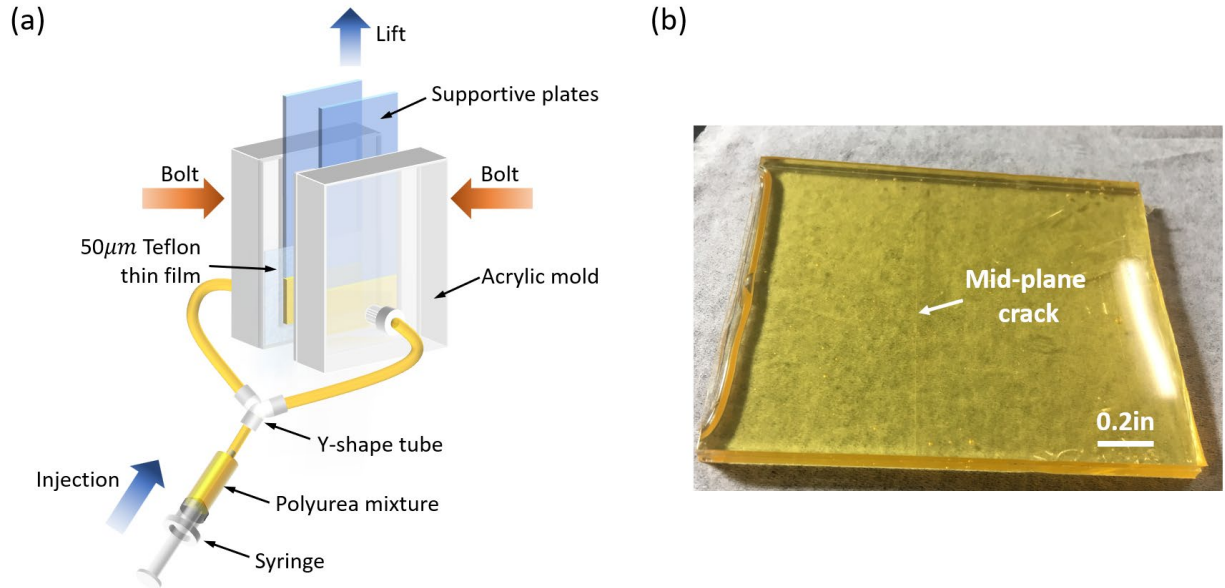

**Figure 7** (a) Injection molding setup to mold polyurea with a sharp mid-plane crack. Polyurea mixture is symmetrically injected by a syringe into two sides of an acrylic mold while a Teflon thin film with $50\mu m$ thickness is clamped between acrylic mold and supportive plates. (b) Molded polyurea specimen with a mid-plane crack.

The thicknesses of the flyer and the specimen were determined to be compatible with the time-distance diagram shown in **Fig. 8(b)**. When the flyer impacts the specimen, compressive waves are generated. After these two waves arrive at the rear surfaces of the flyer and the sample they reflect as unloading (tensile) waves. The stress at the crack plane is released when one of two unloading waves arrives at the crack plane. Therefore, the other unloading wave can open the crack with a traction-free state on the crack faces. For an ideal release wave experiment, the crack will be loaded for a duration of $t_{load} = \frac{2h}{C_l}$, where $h$ is the thickness of the flyer, and $C_l$ is the longitudinal wave speed of polyurea. To ensure the observation time window on the rear surface $(t_{sd}, t_{ed})$ is within the streak camera sweep time window, the ideal thickness of the flyer is selected as $h = 2.22\ mm$, while the ideal thickness of the sample is $H = 2h = 4.44\ mm$. It is noted that $H$ does not have to be exactly $2h$ in the experiments, as we simulate the dynamic loading processes with the experimental specimen dimensions.

## 4.2 Streak-camera triggering-system design

The streak camera system has an intrinsic delay time that depends on the sweep time. If the sweep time is set as $T_{sweep} = 20\ \mu s$, the time difference between triggering the streak camera sweep and the beginning of the sweeping image is $T_{delay} = 13.2\ \mu s$. Therefore, if we trigger the streak camera system at impact, we will miss the observation window. Triggering the system ahead of the impact is crucial to record the fringe evolution within the window of interest. Here, we designed an optical triggering method, where the optical triggering circuit is illustrated as red lines in **Fig. 2**. After the fine alignment between the impact faces of the flyer and the specimen, an acrylic spacer with uniform thickness $d_{sp} = 3.88\ mm$ is placed in between, as shown in **Fig. 8(a)**. Then, a mirror with a sharp edge is mounted inside the fiberglass projectile. The optical circuit is aligned such that the beam reflected from the edge of the mirror is focused onto a photodetector, which triggers a delayer generator (DG645) that controls the streak camera system. The rise time of the photodetector is 100ns, which is negligible in our case. The triggering sequence is illustrated in **Fig. 8(c)**. The time when the streak camera gate is triggered by the photodetector is denoted as $T_0$. Then the streak camera sweep is triggered at $T_1 = 2\ \mu s$. The sweeping image starts at $T_2 = T_1 + T_{delay} = 15.2\ \mu s$, and ends at $T_4 = T_2 + T_{sweep} = 35.2\ \mu s$. Suppose the impact speed of the projectile is $V_0 = 210\ m/s$, the time at impact will be $T_3 = \frac{d_{sp}}{V_0} = 18.48\ \mu s$. In this triggering sequence design, the $10\ \mu s$ window of interest after impact will be centered within the $20\mu s$ sweep time. This triggering design can also accommodate a spacer alignment error up to $1mm$ or an impact speed error up to $\pm 50 m/s$.

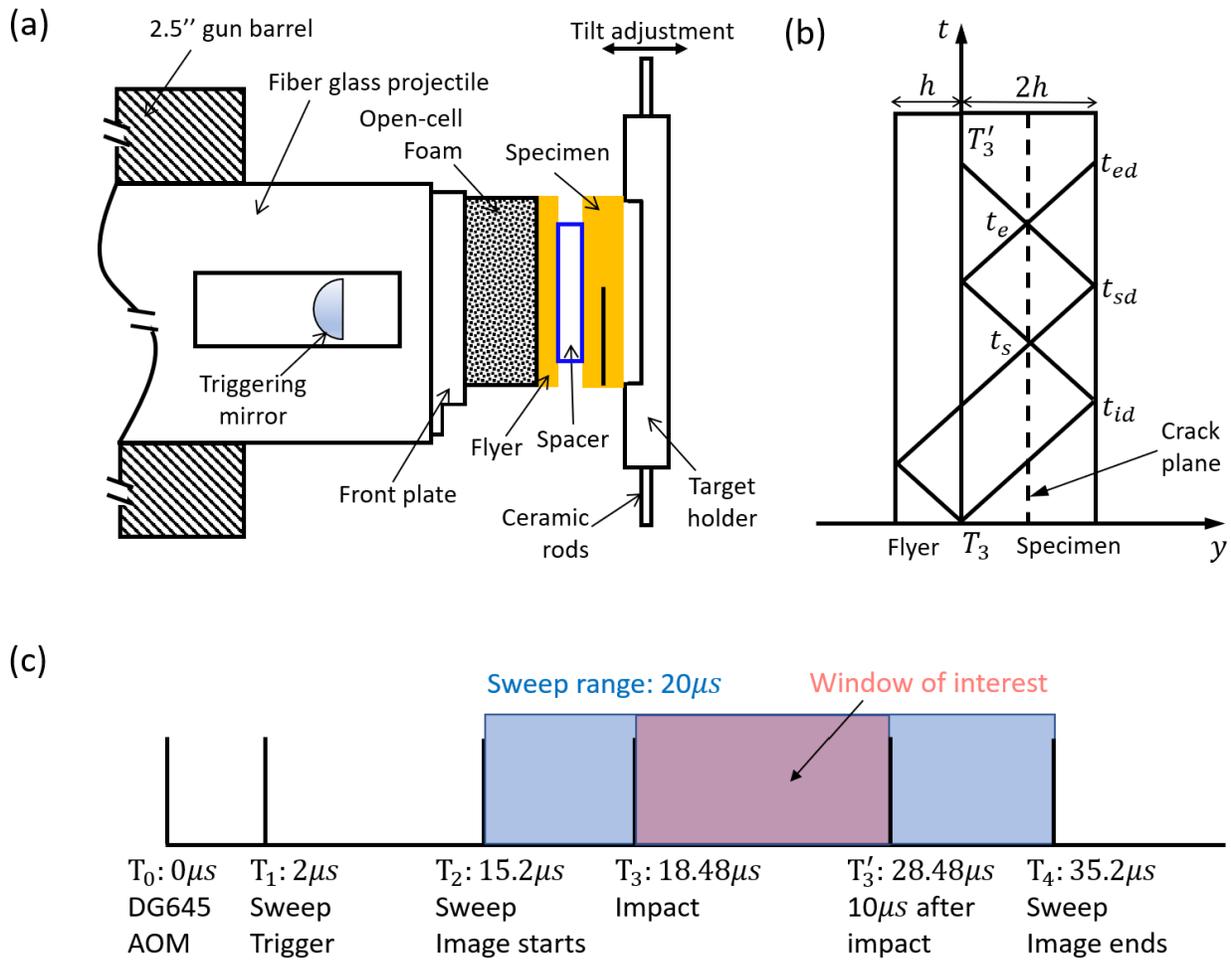

**Figure 8** (a) Delay triggering setup for the streak camera system. (b) Time-distance diagram for an ideal symmetric impact on a specimen with a mid-plane crack. (c) Triggering sequence design.

### 4.3 Plate impact experiments

The plate impact experiments were conducted using a 2.5-inch diameter single-stage gas gun. Detailed experimental procedure for conducting plate impact experiments can be found in Ph.D. theses at Brown University (e.g., Espinosa, 1992; Grunschel, 2009; Kim, 1980; Ravichandran, 1987). Here, key experimental procedures are briefly described below.

To obtain an ideal stress-free boundary on the rear surface of the flyer, a rigid open-cell foam, with 5 $mm$ thickness, was glued between the flyer and the aluminum front plate of the projectile. The aluminum front plate was mounted on a centerless-ground fiberglass projectile tube. An "O" ring at the rear end of the projectile nominally prevented gas from leaking to the front. A Teflon key was used to prevent the rotation of the projectile in the gas gun barrel, which has a keyway. The total mass of the projectile and triggering mirror assembly was approximately 580 g. The polyurea specimen with mid-plane crack was mounted on a Delrin sample holder, which was fixed in an adjustable base by four ceramic rods. An optical alignment technique, developed by Kumar and Clifton, ensures the alignment tilt between the flyer and the sample is less than $2 \times 10^{-5}\ rad$ (Kumar and Clifton, 1977). The projectile was accelerated by a sudden release of nitrogen gas. The velocity of the projectile was measured from the times at which five pairs of wire pins with known distances were shorted by the front plate. The parallelism between the flyer and the projectile was measured from the times at which four tilt pins that were glued at the corners of the specimen were shorted by the aluminum coating on the flyer. The first contact pin triggered the KEYSIGHT oscilloscope for velocity, tilt, and DG645 reference signals. The plate impact experiments were conducted in a vacuum — 60 $mTor$ to 80 $mTor$ — to minimize the air cushion between the flyer and the specimen. Corrugated lead plates were stacked in a catcher tank to decelerate the projectile and the sample after the impact process was completed.

## 5  Results and discussions

We performed three shots with impact speeds of the projectile $V_0$ ranging from 164 m/s to 210m/s. The experimental results are summarized in **Table 1**. Among these three shots, we successfully captured the DL-ISI fringe within the desired window in the third shot. The DL-ISI parameters are listed in **Table 2**. With the selected DL-ISI parameters, the image shearing distance, $\Delta x$ is calculated as 0.111 mm, the initial fringe density, $\rho_{initial}$ is 0.42/mm, and the displacement-gradient sensitivity, $S$ is 2.4 mrad/fringe.

**Table 1** Summary of experimental results. $H$ is the thickness of the specimen with a mid-plane crack, $h$ is the thickness of the flyer, $V_0$ is the impact speed of the projectile, $\theta_t$ is the tilt angle, $\sigma_0$ is the normal stress, and $\Delta c$ is the crack growth length.

| Shots No. | $h$ (mm) | $H$ (mm) | Teflon | $V_0$ (m/s) | $\theta_t$ (mrad) | $\sigma_0$ (MPa) | $\Delta c$ (μm) |
|---|---|---|---|---|---|---|---|
| HJ1901 | 1.98 | 4.44 | N | 164 | 1.70 | 184 | 400 |
| HJ2001 | 2.22 | 4.33 | N | 209 | 2.44 | 234 | 297 |
| HJ2101 | 2.25 | 4.38 | Y | 204 | 3.04 | 229 | 306 |

**Table 2** DL-ISI parameters in HJ2101

| Description | Notation | Value |
|---|---|---|
| Laser wavelength | $\lambda$ | 532 nm |
| Laser-illumination window length | $L_b$ | 10 mm |
| Laser-illumination window width | $W_b$ | 2 mm |
| Streak-camera slit width | $W_{sk}$ | 100 μm |
| Refractive index of glass plate | $n_g$ | 1.52 |
| Beam incident angle | $\theta_{in}$ | $45^0$ |
| Glass thickness | $h_g$ | 150 μm |
| Image plane distance | $f_{img}$ | 500 mm |
| Image-shear distance | $\Delta x$ | 0.111 mm |
| The number of initial fringes | $N_{inif}$ | 4.2 |
| Initial fringe density | $\rho_{initial}$ | 0.42 /mm |
| DL-ISI fringe sensitivity | $S$ | 2.4 mrad/fringe |

## 5.1 DL-ISI fringes

The raw experimental fringe image from the streak camera system for Shot 2101 is shown in **Fig. 9**. The horizontal axis represents the spatial information along a line perpendicular to the crack plane at the rear surface with $x = 0$ aligned at the initial mid-crack tip position, while the vertical axis represents the temporal information with $t = 0 \mu s$ denoting the time at impact.

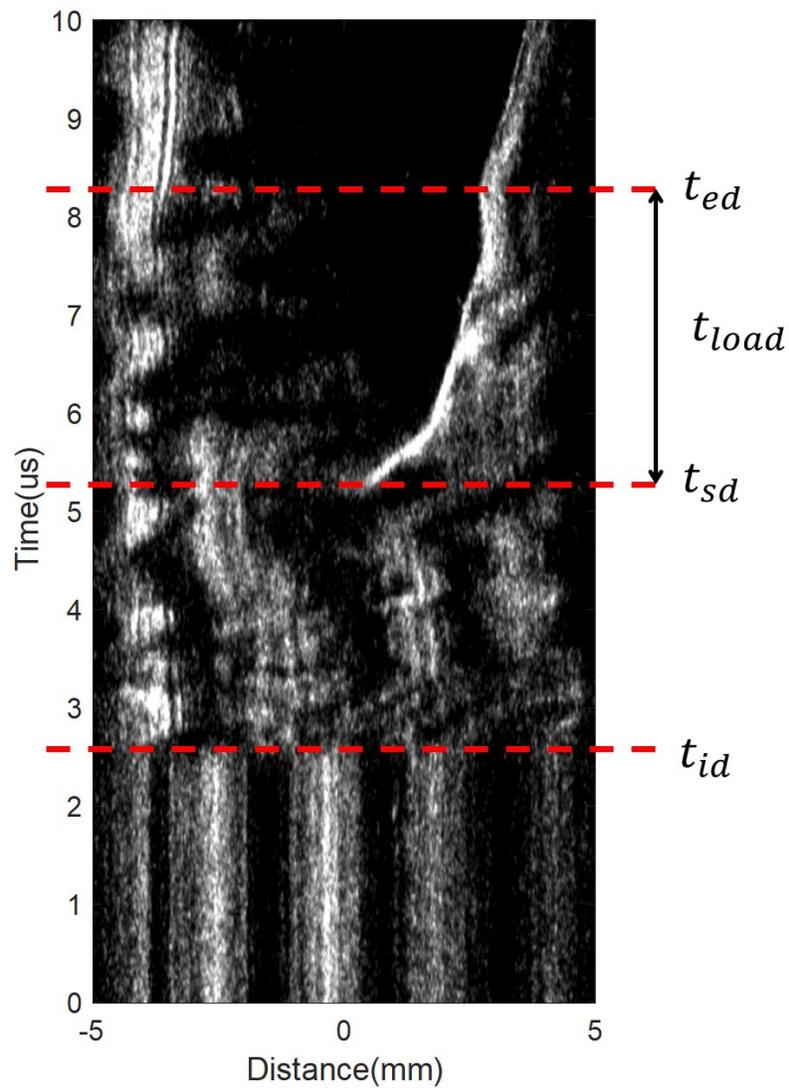

**Figure 9** Raw experimental DL-ISI fringe image measured from Shot HJ2101. The time markers correspond to those in **Fig. 8(b).**

To correct the image noise and inhomogeneous illumination issues of the streak camera system, the raw experimental fringe image was filtered by the $3 \times 3$ low pass filters and then passed through an intensity-normalization filter. The resultant fringe image is shown in **Fig. 10(a)**. To understand the general feature of the experimental fringes, a benchmark FEM simulation with a perfectly bonded interface condition was performed. The computational fringes are plotted in **Fig. 10(b)**. When $t < t_{id} = 2.7 \ \mu s$, the compressive wave has not arrived at the rear surface. Therefore, a slit image of the initial fringes sweeps temporally to exhibit initial DL-ISI fringes. From **Eq. 8**, the number of initial DL-ISI fringes is estimated to be $N_{inif} = 4.2$, which is consistent with the experimental observation. The 3.04mrad tilt angle between the flyer and the specimen generated a tilt of the compressive wavefront of $1.7^0$, which led to a shift of initial fringes, as observed in both experimental and FEM fringe images. During the crack loading time interval ($t_{sd} < t < t_{ed}$), the tensile wave reflected from the crack tip arrived at the rear surface. Thus, the fringe density varied according to the change of displacement gradient due to crack opening.

As we can see in **Fig. 10(a)**, the experimental fringe captured the correct crack loading period. However, fringes were displayed only away from the crack-tip trajectory projected on specimen's rear surface, missing fringes near the projected trajectory. There are two potential reasons for this phenomenon. First, the large surface-displacement gradient near the projected trajectory during the tensile wave loading period could cause the reflected laser beam to miss the collimating lens (SL2 in **Fig. 2**). Therefore, the laser beam position at $x$ in the following relation,

$$\left| x - 2 \frac{\partial u(x,t)}{\partial x} \cdot f_{len} \right| > \frac{d_{len}}{2}, \tag{11}$$

could not be observed in the experiment. Here, $d_{len} = 50 \ mm$ is the diameter of the collimation lens and $f_{len} = 500 \ mm$ is the focal length. The FEM fringes which reflected this surface-warping effect is plotted in **Fig. 10(c)**. Indeed, most of the fringe information from the specimen's rear surface nearest to the projected crack-tip trajectory is missing due to the large surface-displacement gradient. So far, we have considered only the surface warping effect out of the first collimation lens. However, in the experiment, some reflected laser beams could also move out of the range of the second collimation lens (SL3 in **Fig. 2**) or other optics. Therefore, the missing portion of the fringe can be expected to be larger in the experiment, than in the simulation shown in **Fig. 10(c)**.

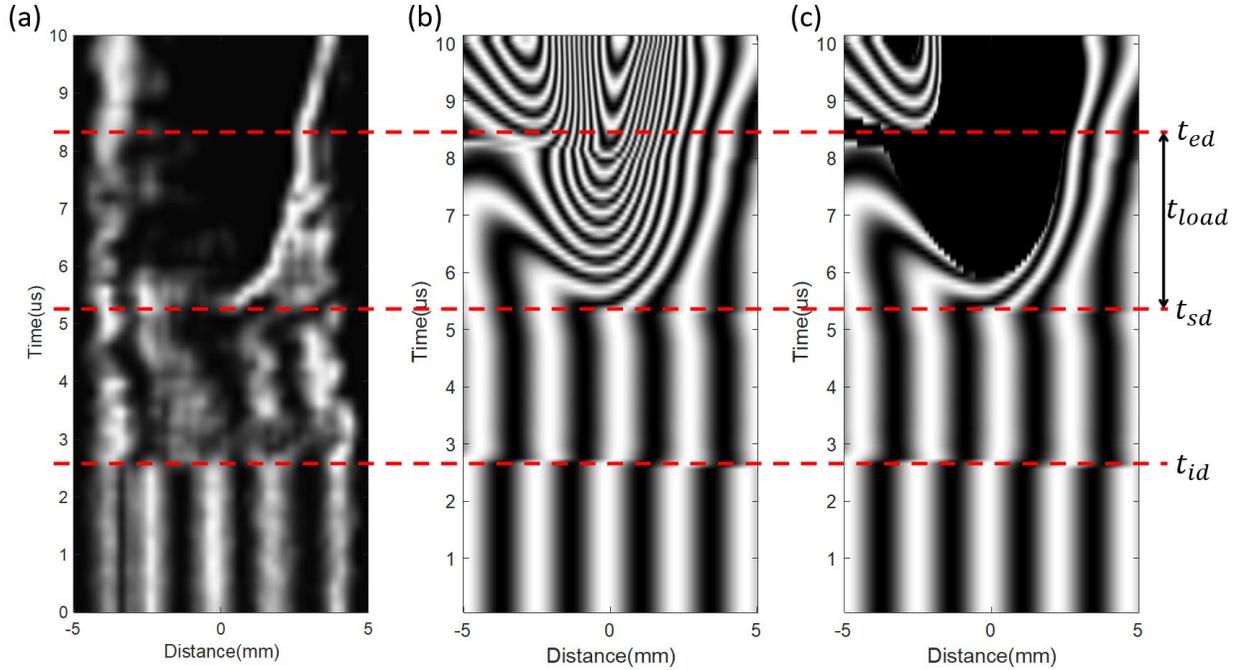

**Figure 10** (a) Smoothed experimental DL-ISI fringe image; (b) Computational DL-ISI fringe image from FEM with a perfectly-bonded interface; (c) Computational DL-ISI fringe image which reflects surface-warping effect.

Second, from FEM, the compressive strain at the rear surface ahead of the projected crack-tip location is found to be ~6%. This compressive strain is larger than the critical buckling strain of the 200 nm aluminum coating on the polyurea substrate, $\varepsilon_{bc} = \frac{1}{4}\left(\frac{3E_{PU}}{E_{Al}}\right)^{2/3}$ ~4%, which could result in a buckling wavelength, $\lambda_{bc} = 2\pi h_{Al}\left(\frac{E_{Al}}{3E_{PU}}\right)^{\frac{1}{3}}$ ~2.3 $\mu m$. The buckled aluminum coating could diffract the lights to form the image with missing fringes, as observed in **Fig. 9**. Delamination or cracking of the aluminum coating could also be responsible for the incomplete fringe image. Unlike the traditional dynamic crack-tip region imaging employed in the method of caustics (Ravichandar and Knauss, 1984) or coherent gradient sensing (Tippur et al., 1991) taken at discrete frozen times, the DL-ISI data on a finite domain of the specimen's rear surface provides convoluted information on the crack opening history. Although we miss a zone of large displacement gradient on the t-x plane of the rear surface, that does not mean that we miss the information from the crack-tip region. We believe that with fringe inpainting our cohesive-zone functional form is well suited to regularizing the inversion process of getting the two cohesive zone parameters as shown in the next section. At this point, the most reassuring information set of the deep-learning assisted inversion is the post-mortem measurement of the cohesive-zone size, opening displacement, and crack propagation length, as is presented in **Section 5.3**.

## 5.2 Physics-based DL-ISI fringe inpainting with Conditional Generative Adversarial Nets (cGAN)

As discussed in the previous section, experimental fringe information was missing when the reflected wave from the crack-plane arrived at the rear surface of the polyurea specimen. Nevertheless, a distinct fringe near the boundary of the fringe-missing-zone was successfully recorded in the experiment. The characteristic shape of the fringe is consistent with the FEM fringe. Therefore, if we could inversely obtain the missing fringes based on this boundary fringe, we could understand the crack propagation history and hence measure the dynamic fracture toughness of polyurea.

This physics-based fringe inpainting is possible because of the recent development of Conditional Generative Adversarial Nets (cGAN) (Mirza and Osindero, 2014). Although the original idea of a GAN (Goodfellow et al., 2014) is used for unsupervised learning without giving specific labels, a cGAN can generate new data based on a given condition. For example, in our case, we would like to train a cGAN model to generate a full fringe image from the given condition of a partial fringe image. Since its original development, cGAN has been successfully used in image-to-image transitions like image semantic segmentation (Rezaei et al., 2017) and image inpainting (Isola et al., 2016). As for its application in Solid Mechanics, cGAN can inversely identify the material modulus distribution from the given strain distribution images (Ni and Gao, 2021) or predict strain and stress distributions for complex composites (Yang et al., 2021a; Yang et al., 2021b). However, most current applications of cGAN in engineering have been limited to computational simulation datasets. Here, we apply cGAN to the physics-based experimental fringe inpainting. To our knowledge, this is the first application of cGAN used for real experimental data cleaning.

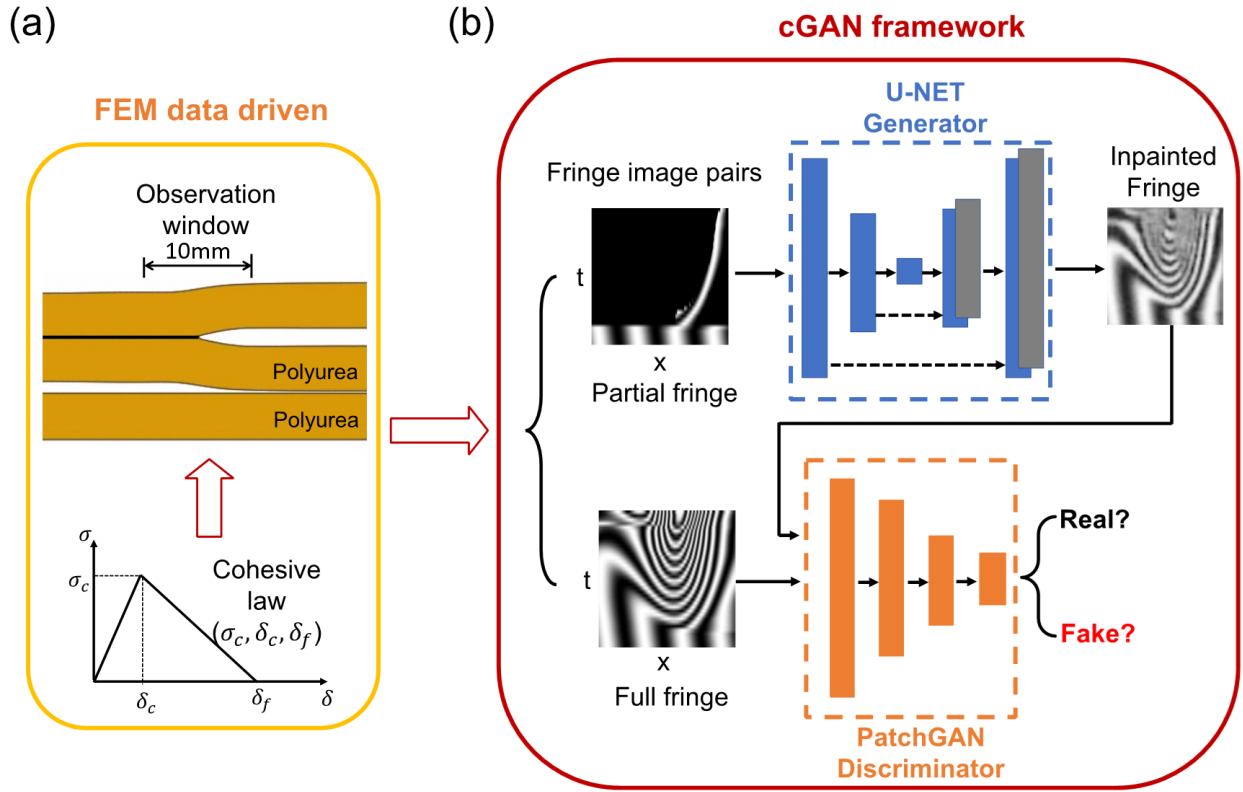

**Figure 11** Schematics of the cGAN model consisted of a generator with U-NET architecture and a discriminator with PatchGAN architecture for the physics-based DL-ISI fringe inpainting.

As shown in **Fig. 11**, the typical cGAN model consists of a generator with U-NET architecture (Ronneberger et al., 2015) and a discriminator with PatchGAN architecture (Isola et al., 2016). The U-NET with skip-connection techniques is used to generate the full fringe images from the input of the given missing fringe images, while the PatchGAN evaluates the generated fringe images by comparing with the input images patch by patch. During training, the generator generates new fringe images as authentic as the ground-true fringe images while the discriminator learns to perform better to distinguish the new image from the ground truth. The performance of these two adversarial networks improves until reaching a Nash equilibrium when the generated fringe image is too real to be distinguished by the discriminator. For the detailed training loss function definition for the cGAN model, readers can refer to the original cGAN paper (Mirza and Osindero, 2014).

The cGAN model was trained on a dataset with 961 pairs of full and partial fringe images obtained from FEM with different 2-parameter ($\sigma_c$, $\delta_f$) bilinear cohesive laws. To produce a training dataset as consistent as the experimental fringe, the displacement gradient at the rear surface that satisfies **Eq. 11** was removed to produce the partial fringe images as input images. Furthermore, the left portion of the computational fringe images was removed since the aluminum coating on the left side of the rear surface was damaged. Examples of input and ground-true fringe images are shown in **Fig. 11(b)**. The total dataset with 961 fringe images was randomly split into 90% train dataset and 10% test dataset. The training fringe images were rescaled to $128 \times 128$ pixels. The normalized L2 error is used as the cGAN model performance matrices as,

$$E_{L2} = \sqrt{\frac{\sum_{i,j}\left(I_{ij}^{pre} - I_{ij}^{true}\right)^2}{\sum_{i,j}\left(I_{ij}^{true}\right)^2}} \times 100\%, \quad (12)$$

where $I_{ij}^{true}$ and $I_{ij}^{pre}$ are the intensities at pixel $i, j$. The detailed model architectures for generator and discriminator are listed in **Table B.1** and **B.2** in **Appendix B**. The training was performed on the open-source platform TensorFlow V2.5 (Abadi et al., 2016). The Adam optimizer (Kingma and Ba, 2014) with a learning rate of 0.0002 was used to optimize the cGAN model. The batch size was set as 32 in our model.

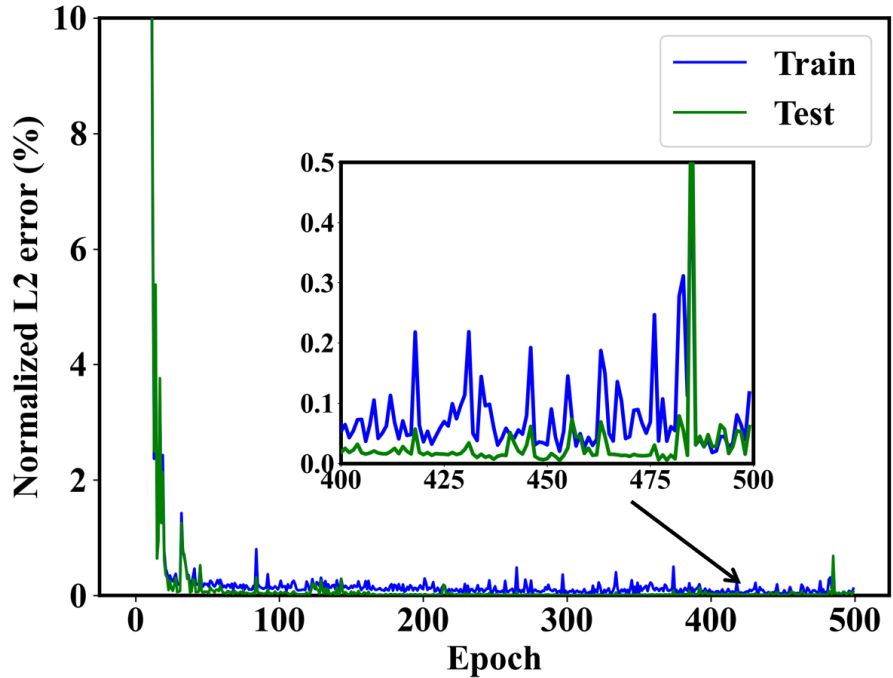

**Figure 12** Training performance of the cGAN model as a function of train epoch.

The normalized L2 errors of train and test datasets as a function of train epochs are shown in **Fig. 12**. The total training time for 500 epochs on a 32-core CPU is approximately 8 hours. As shown, both errors rapidly reduced to 2% after 10 epochs and converged to 0.05% after 100 epochs. The errors of train and test datasets are similar, so no overfitting is observed. Because of the image-to-image mapping capability of the cGAN model, training 200 epochs on the current cGAN model with a 961 FEM dataset is sufficient to make a prediction with an error of 0.05%. To visualize the cGAN model performance, the predicted fringe images from two randomly selected input fringe images from the test datasets after 200 epochs are shown in **Fig. 13**. The pixel-wise L2 errors between prediction and ground-truth are also plotted. As one can see, the majority of pixels have an error less than 0.2%, and the pixels with the largest errors are at the boundary of each fringe. These results illustrate the accurate performance of pixel-wise fringe inpainting from the cGAN model. Besides its high accuracy, the well-trained cGAN model is also efficient in generating a new fringe image. For example, it takes approximately 5 *ms* on a personal computer for the cGAN model to generate a new fringe image for a new input partial fringe.

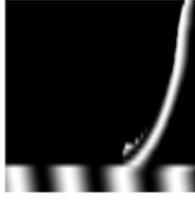

**Figure 13** Input fringe images, ground-truth fringe images, predicted fringe images, and the normalized L2 errors of two randomly selected input images in the test dataset after 200 epochs of training.

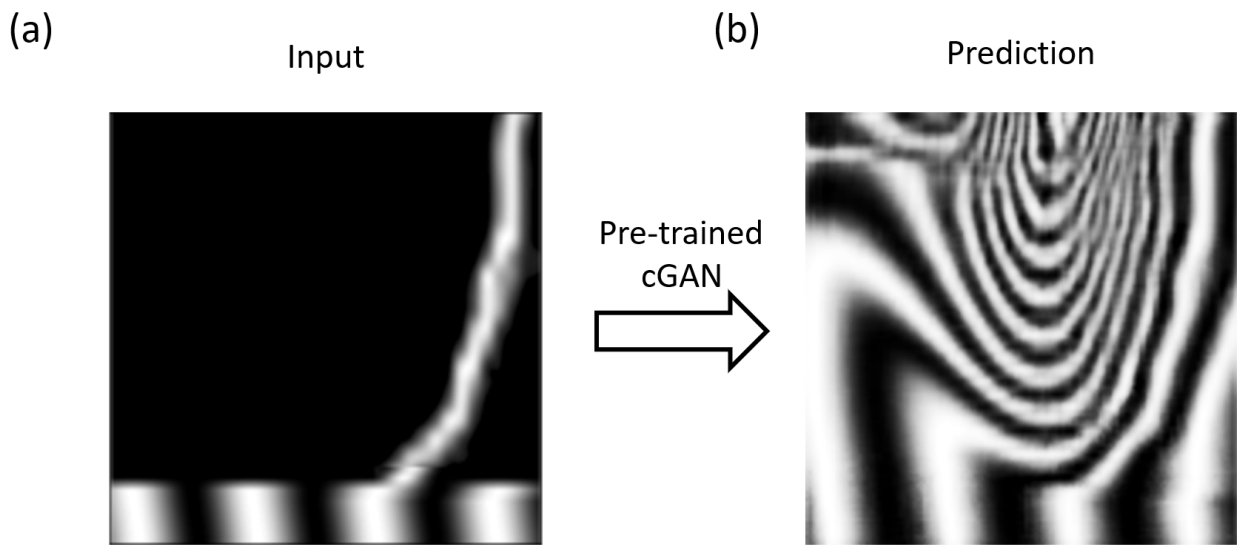

**Figure 14** Fringe inpainting of experimental fringe image with initial fringe padding from pre-trained cGAN with 200 epochs.

After training the cGAN model on the FEM fringe dataset for 200 epochs, the filtered experimental DL-ISI fringe image (**Fig. 10(a)**) was cropped with a maximum correlation window with FEM datasets and rescaled to 128 × 128 pixels as the input image to generate the inpainted fringe image. Compared with the FEM fringe image, the experimental fringe image does not have clear initial fringes due to the surface tilt effect. To be consistent with the FEM train dataset, we prepared an experimental fringe image input with initial fringe padding from FEM as well. Then, the padded fringe image was processed with histogram equalization with the FEM inputs. The processed experimental fringe is shown in **Fig. 14(a)**. The inpainted fringe images from the input images are shown in **Fig. 14(b)**. The missing fringes were able to be inpainted inversely from the characteristic boundary fringe measured in the experiment. It is worth noting that the generated fringe image becomes locally blurred and zigzag, especially where the fringe density is high, compared with the FEM dataset. The characteristic fringe is not as smooth as in the FEM dataset, resulting in noise in the predicted fringes. Nevertheless, the pre-trained cGAN model with a computational dataset is capable of generating a physics-based full-field fringe image from the experimental data.

## 5.3 Cohesive parameters and dynamic fracture toughness of polyurea

Since we obtained the full-field fringe image during the crack loading process from the cGAN model, the cohesive parameters could ideally be determined through the deep learning framework proposed in **Section 2**. However, as discussed in the previous subsection, the inpainted fringe is locally zigzag and noisier compared with the FEM dataset, reducing the accuracy of CNN predictions. Here, we calculate the correlations between the inpainted fringe and each FEM fringe in the FEM dataset in order to statistically determine the cohesive parameters as well as the fracture toughness. The CNN prediction is also then included for comparison. The correlation coefficient is defined as,

$$cor = \frac{\sum_{i,j}(I_{ij}^{exp} - \bar{I}^{exp})(I_{ij}^{FEM} - \bar{I}^{FEM})}{\sqrt{\sum_{i,j}(I_{ij}^{exp} - \bar{I}^{exp})^2 \sum_{i,j}(I_{ij}^{FEM} - \bar{I}^{FEM})^2}}, \quad (13)$$

where $I_{ij}^{exp}$ and $I_{ij}^{FEM}$ are the intensities at pixel $i,j$ for experimental and FEM fringe image, respectively. $\bar{I}^{exp}$ and $\bar{I}^{FEM}$ are the mean values. The correlation coefficients, plotted in **Fig. 15**, are in the range of 0.35~0.89. To determine the cohesive parameters as well as the standard deviations, the correlation coefficients with $cor > 0.75$ are fitted into a bivariate normal distribution. The prediction results are summarized in **Table 3**. The predicted cohesive parameters with error bars are marked in yellow in **Fig. 15**. The dynamic fracture toughness of polyurea under an extremely high rate of crack-tip loading $\dot{K} \sim 10^7 \, MPa\sqrt{m}/s$ is determined as $G_c = 12.1 \, kJ/m^2$ with a standard deviation of $1.10 \, kJ/m^2$. The estimation of dynamic fracture toughness based on the linear elastic fracture mechanics (LEFM) theory is $G_c^{LEFM} = 13.4 \, kJ/m^2$ (Details in **Appendix C)**, which overestimates by ~11%. The CNN prediction based on 7 independent trainings is summarized in **Table 3** and marked in green in **Fig. 15**.

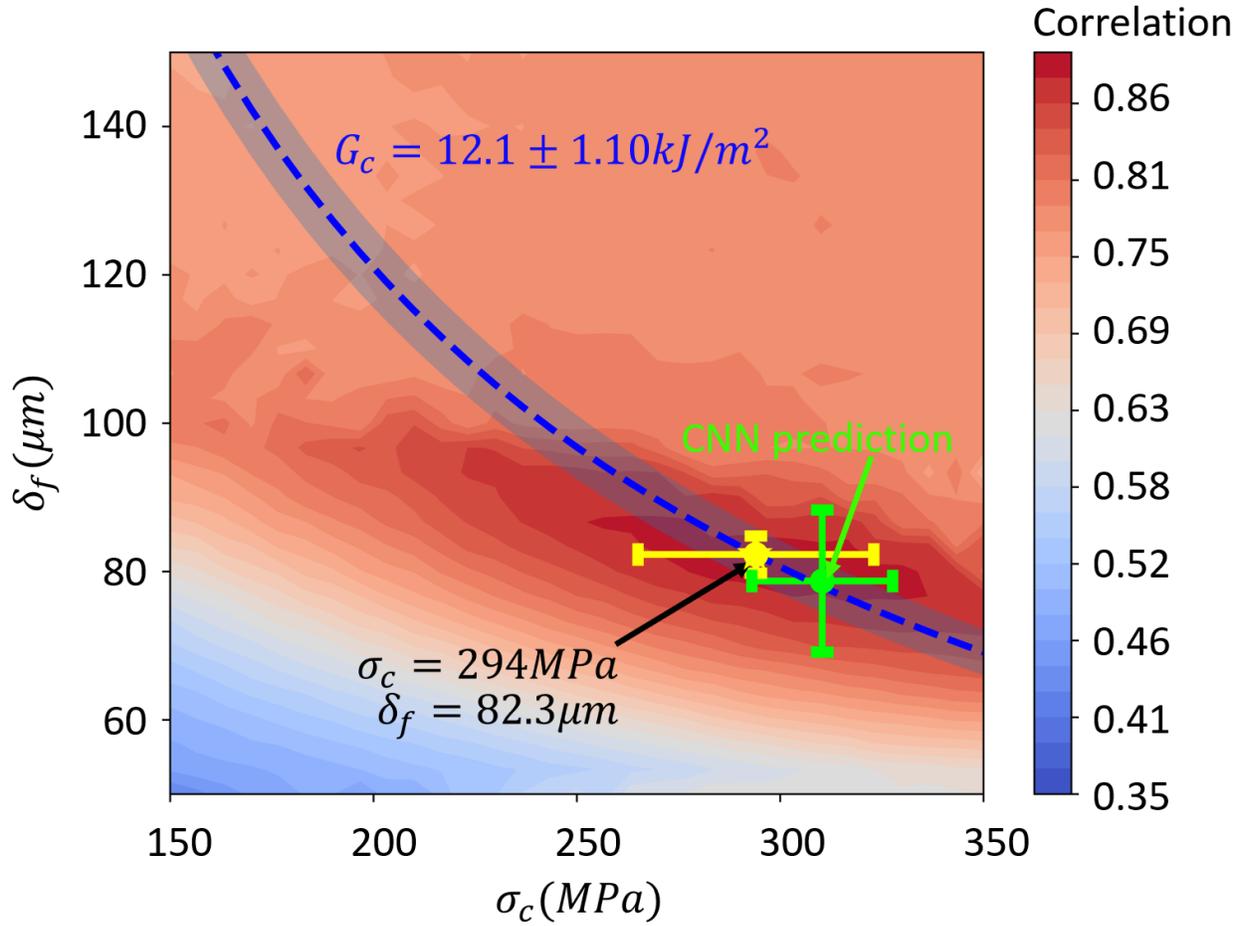

**Figure 15** Correlation coefficients contour between the experimental fringe and 961 FEM fringes in the dataset. The cohesive parameters with error bars determined from bivariate normal distribution are marked in yellow. The dynamic fracture toughness is marked in blue dotted line with shaded error bar. The CNN predictions with error bars are marked in green.

**Table 3** Cohesive parameters predictions from the correlation and CNN methods

| Correlation | | | CNN | | |
|---|---|---|---|---|---|
| $\sigma_c(Mpa)$ | $\delta_f(\mu m)$ | $G_c(kJ/m^2)$ | $\sigma_c(MPa)$ | $\delta_f(\mu m)$ | $G_c(kJ/m^2)$ |
| 294 | 82.3 | 12.1 | 310 | 78.7 | 12.2 |
| $\pm 29$ | $\pm 2.5$ | $\pm 1.1$ | $\pm 17$ | $\pm 9.6$ | $\pm 1.7$ |

As we can see, the CNN makes a reasonable prediction of the fracture toughness with only a ~1% difference with the correlation method. However, more significant prediction discrepancies have been made in the individual parameters. Furthermore, CNN has a more significant prediction variance than the correlation methods. This difference in variance is due to the difference in fidelity between our train dataset and the experimental test dataset, such that our training dataset is the 2D FEM dataset with low fidelity. However, the experimental inputs have high fidelity. Therefore, constructing the correlation between these two datasets plays a crucial role in the multi-fidelity modeling method (Fernández-Godino et al., 2016). To this end, Meng and Karniadakis developed a multi-fidelity deep neural network (MFNN) that can efficiently train the datasets with different levels of fidelity (Meng and Karniadakis, 2020). For example, compared with the single-fidelity training, MFNN is capable of increasing the prediction accuracy on the material properties in the instrumented indentation when trained on a multi-fidelity dataset with both FEM and experiments (Lu et al., 2020). Therefore, as more experimental fringes are collected in future experiments, one could use MFNN to train the FEM fringes as well as the previous experimental fringes together to increase the accuracy of the predicted cohesive parameters. Except for the error source from the multi-fidelity, other error sources on the predictions could come from the sensitivity of the DL-ISI fringe, cGAN, and CNN models. These errors may not be independent; thus, distinguishing the errors among these different sources may not be feasible.

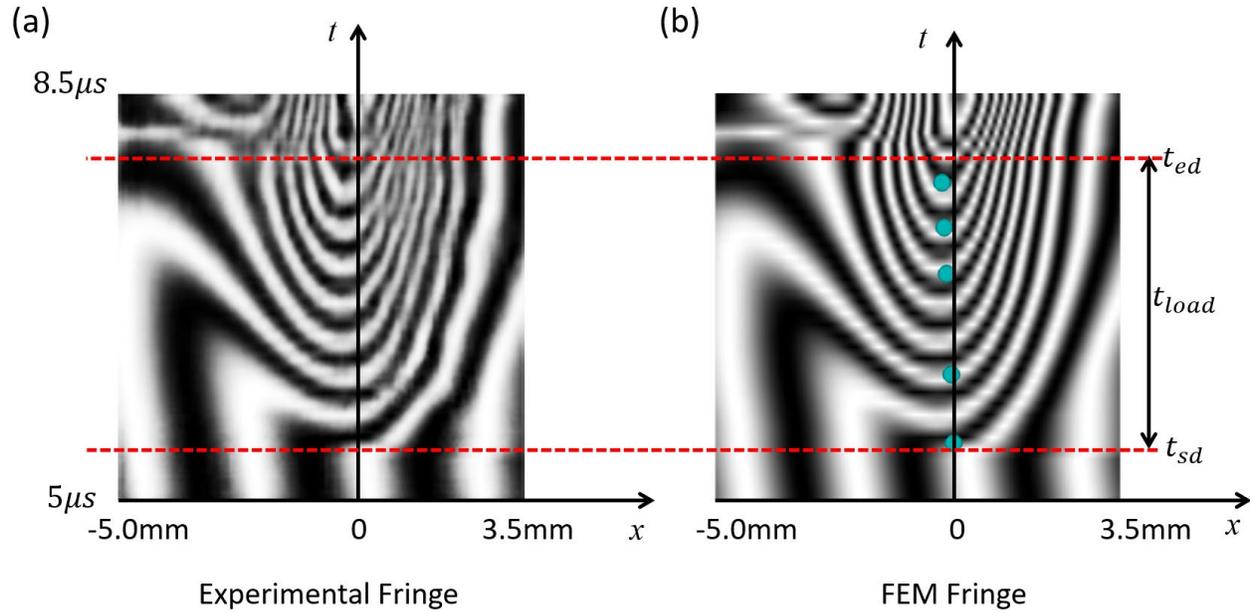

**Figure 16** (a) Inpainted experimental fringe; (b) FEM fringe generated from FEM simulation with predicted cohesive parameters of $\sigma_c = 294\ MPa$ and $\delta_f = 82.3\ \mu m$. The crack tip positions are labeled in green dots.

When the accurate cohesive parameters are determined, the crack growth history can be readily analyzed from the FEM. The FEM fringes generated from the simulation with the predicted cohesive parameters are plotted together with the experimental fringes in **Fig. 16**. Two fringe images are very similar, with a correlation coefficient of 0.881. The crack-tip positions are labeled by green markers on the FEM fringe image. The maximum crack growth length during the first crack loading period is $250\ \mu m$. To understand the crack growth process, we plotted the three crack advance distances, $a_{0.1}$, $a_c$, and $a_f$ in **Fig. 17(b)**, when the separations between crack surfaces reach $0.1\delta_c$, $\delta_c$, and $\delta_f$ as defined in **Fig. 1(b)** and shown in **Fig. 17(a)**. The result shows that $a_c$ increases to $350\ \mu m$ while $a_f$ increases to $250\ \mu m$ during the first period of crack loading. The fractography of the crack plane after the impact was obtained from optical microscopy and shown in **Fig. 17(c)**. The crack-tip positions with $a_c$ and $a_f$ are labeled in the fractography. We found the crack front of $a_c$ is straight while the crack front of $a_f$ is relatively wavy with the average wavelength of 100~200 $\mu m$ and the average crack advance distance of 200~250 $\mu m$. These crack growth distances are consistent with FEM results during the first loading period. A wavy crack front with micron-scale wavelength could be caused by the local material inhomogeneity. Clusters with the size of hundred nanometers were observed in bulk polyurea from our high-resolution AFM tapping-mode morphology images (Kim et al., 2020). However, in the micron-scale, the polyurea sample is considered as homogenous, and linear-elastic crack-front growth instability is not expected (Gao and Rice, 1989; Rice et al., 1994). Therefore, in the experiment, the crack front when the surface separation reached $\delta_c$ is straight even though the initial stationary crack front is slightly wavy. However, the $a_f$ crack front could become wavy as the crack-face traction softens for $\delta_c < \delta < \delta_f$. In addition, during the unloading process, the fracture wake zone can be partially healed due to the self-healing capability of polyurea, leaving local craze zone residuals on the crack plane. These residuals could enhance the apparent waviness of the post-mortem $a_f$ crack front. Nevertheless, the 3D crack advance and self-healing processes were not considered in our FEM simulations for computational efficiency. The cohesive strength $\sigma_c$ and the toughness of polyurea $G_c$ under the crack loading rate are displayed on an Ashby diagram (Ashby, 2012) in **Fig. 18**. Under the dynamic loading, the cohesive strength of polyurea is found in the cohesive-property range of composites, which is far beyond the strength of any other polymers and rubbers.

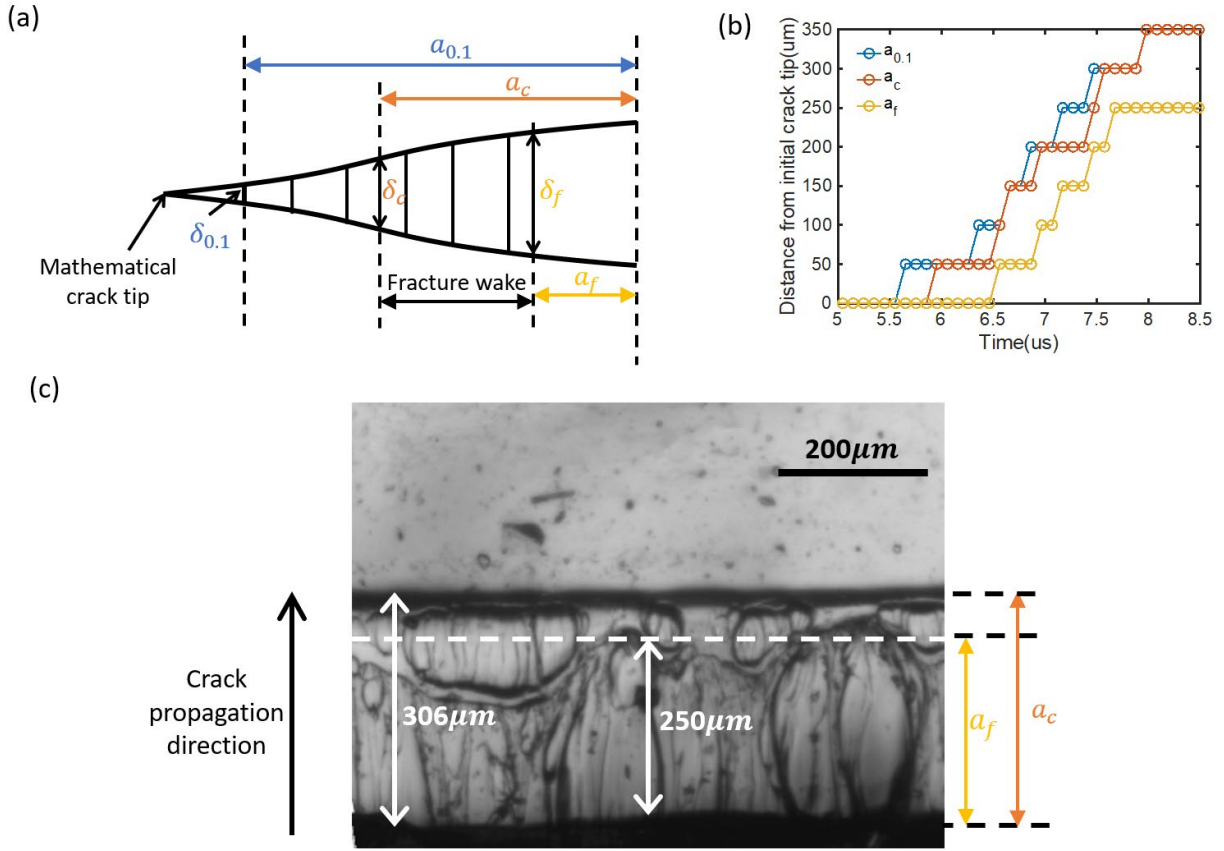

**Figure 17** (a) The schematics of crack process zone. (b) The crack advance distances of $a_{0.1}$, $a_c$, and $a_f$ as a function of time in the FEM simulations. (c) Fractography for Shots HJ2101.

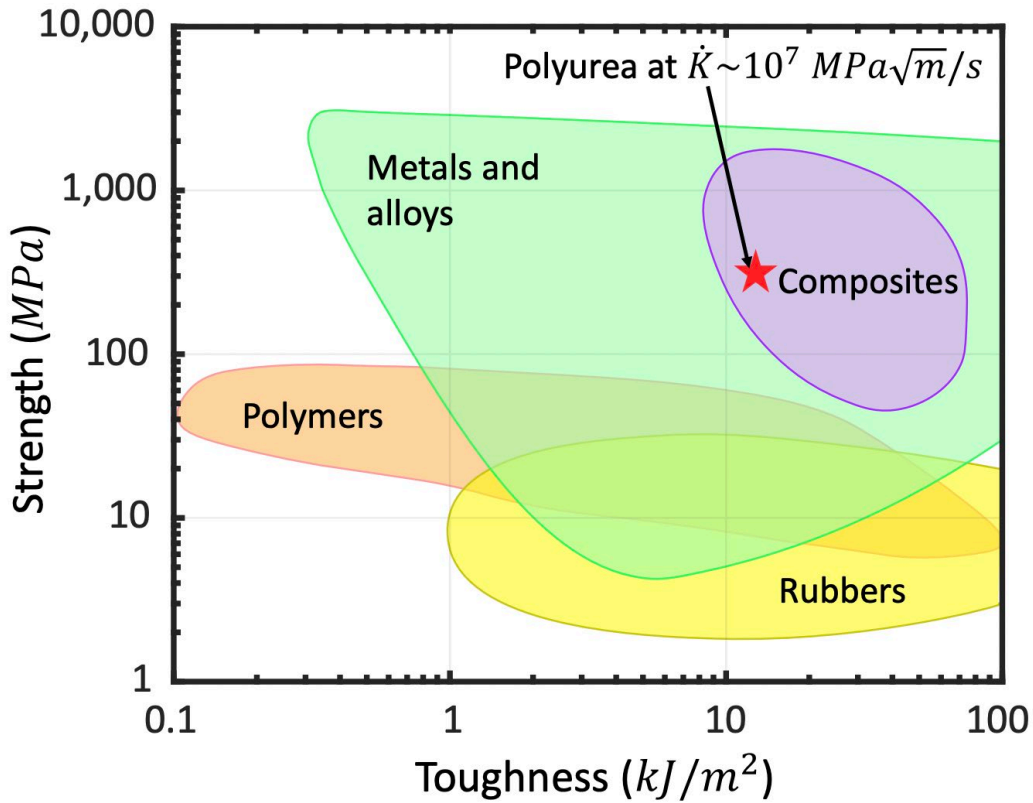

**Figure 18** High dynamic fracture toughness of polyurea at a high crack-tip loading rate plotted on an Ashby diagram (Ashby, 2012).

# 6 Conclusion

In this paper, detailed dynamic cohesive properties (DCPs) beyond the dynamic toughness of polyurea PU1000 were successfully measured from a big-data-generating experiment with our newly introduced DL-ISI, applying plane-wave loading on a mid-plane-cracked specimen with plate impact. Unlike the conventional high-speed photography of spatial optical fringes employed in dynamic-fracture testing with complex 3D free-surface effects, the DL-ISI generates optical fringes of dynamic-fracture processes under plane-strain conditions on a time-space coordinate. The imaging system of the DL-ISI missed some fringes of large displacement gradient;

nevertheless, the missing fringes could be reconstructed by a state-of-the-art cGAN model pre-trained with a corresponding FEM-simulation dataset. Then, a CNN and a correlation method were applied to the fully reconstructed fringes to get the dynamic fracture toughness, $12.1 \pm 0.05 \; kJ/m^2$, cohesive strength, $302 \pm 8.2 \; MPa$, and maximum cohesive separation, $80.5 \pm 1.8 \; \mu m$, under a very high crack-tip loading rate, $\dot{K} \sim 10^7 \; MPa\sqrt{m}/s$. The DCPs were delineated from polyurea's dynamic nonlinear bulk behavior by employing physics-based instantaneous finite-deformation constitutive relations of (Clifton and Jiao, 2015) in the finite element simulations for the deep-learning analyses. The dynamic cohesive strength, $\sigma_c = 302 \pm 8.2 \; MPa$, is found to be nearly three times higher than the dynamic-failure-initiation stress ~105 MPa under a symmetric impact with the same impact speed (Clifton and Jiao, 2015). The dynamic cohesive separation is also very large under such a high rate of loading. The results show that the high dynamic fracture toughness originates from both high dynamic cohesive strength and high ductility of the dynamic cohesive separation. The high dynamic cohesive strength is believed to come from the dynamic load-sharing characteristics of polyurea's hard and soft nanophases and the high dynamic ductility from the fragmentation and self-healing characteristics of the rod-shaped hard nanophase embedded in the soft matrix. The underlying strengthening mechanisms were investigated through molecular-level *in-situ* experiments and mesoscale simulations, and reported in a sequel (Jin et al., 2022b). These experimental results fill a gap in the current understanding of bicontinuously nanostructured copolymer's cooperative-failure strength under extreme local conditions near the crack tip. We also believe the framework of this *big-data-generating experiment* could steer the future direction on experimental mechanics, which combines the newly introduced high-throughput experimental techniques with state-of-the-art machine learning algorithms for the next-level material design and characterization.

## Acknowledgments

The authors gratefully acknowledge the supports provided by U.S. Office of Naval Research (Grant N00014-18-1-2513). Helpful discussions with Mr. Enrui Zhang on deep learning section are acknowledged.

## CRediT author statement

**Hanxun Jin:** Conceptualization, Methodology, Software, Data Curation, Validation, Writing - Original Draft

**Tong Jiao:** Software, Data Curation

**Rodney J. Clifton:** Writing - Review&Editing, Supervision, Project administration

**Kyung-Suk Kim:** Conceptualization, Methodology, Writing – Review & Editing, Supervision, Project administration

# Appendix A

# DL-ISI fringe sensitivity on the detailed cohesive laws

| Cohesive law | Bilinear | Trapezoid | Exponential |
|---|---|---|---|
| Schematics | 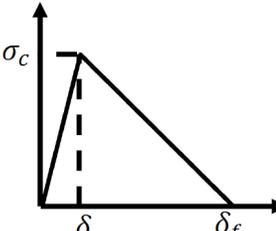 | 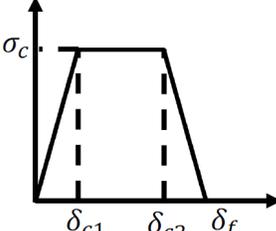 | 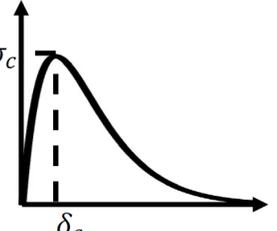 |
| Cohesive parameters | $\sigma_c = 300 MPa$<br>$\delta_f = 80\mu m$<br>$\delta_c = 0.2\delta_f$ | $\sigma_c = 300 MPa$<br>$\delta_f = 53\mu m$<br>$\delta_{c1} = 0.25\delta_f$<br>$\delta_{c2} = 0.75\delta_f$ | $\sigma_c = 300 MPa$<br>$\delta_c = 15\mu m$ |
| Fracture toughness | $G_c = 1.20 kJ/m^2$ | | |
| DL-ISI fringes | 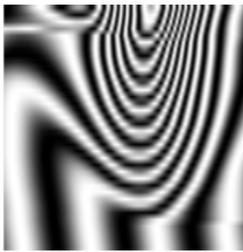 | 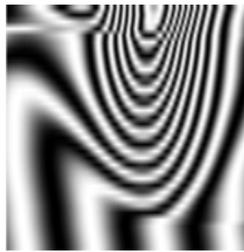 | 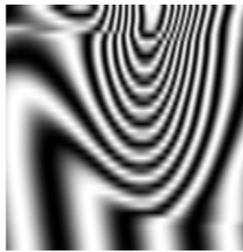 |
| Correlation | 1.0 | 0.9909 | 0.9807 |
| Wall time | 10.8s | 9.8s | 18.3s |

**Figure A.1** DL-ISI fringe images from FEM simulations with 3 different cohesive laws: bilinear, trapezoid, and exponential laws with the same fracture energy.

# Appendix B

# Detailed cGAN model

**Table B.1.** Detailed U-Net Generator architecture.

| Layer | Layer structure | Output shape |
|---|---|---|
| 1 | Input layer | (128, 128, 1) |
| 2 | Conv2D + LeakyReLU(alpha=0.2) | (64, 64, 64) |
| 3 | Conv2D + BatchNomalization + LeakyReLU(alpha=0.2) | (32, 32, 128) |
| 4 | Conv2D + BatchNomalization + LeakyReLU(alpha=0.2) | (16, 16, 256) |
| 5 | Conv2D + BatchNomalization + LeakyReLU(alpha=0.2) | (8, 8, 512) |
| 6 | Conv2D + BatchNomalization + LeakyReLU(alpha=0.2) | (4, 4, 512) |
| 7 | Conv2D + BatchNomalization + LeakyReLU(alpha=0.2) | (2, 2, 512) |
| 8 | Conv2D + BatchNomalization + LeakyReLU(alpha=0.2) | (1, 1, 512) |
| 9 | UpSampling2D + Conv2D + BatchNomalization | (2, 2, 1024) |
| 10 | UpSampling2D + Conv2D +BatchNomalization | (4, 4, 1024) |
| 11 | UpSampling2D + Conv2D +BatchNomalization | (8, 8, 1024) |
| 12 | UpSampling2D + Conv2D +BatchNomalization | (16, 16, 512) |
| 13 | UpSampling2D + Conv2D +BatchNomalization | (32, 32, 256) |
| 14 | UpSampling2D + Conv2D +BatchNomalization | (64, 64, 128) |
| 15 | Output layer | (128, 128, 1) |

**Table B.2.** Detailed PatchGAN Discriminator architecture.

| Layer | Layer structure | Output shape |
|---|---|---|
| 1 | Concatenate layer | (128, 128, 2) |
| 2 | Conv2D + LeakyReLU (alpha=0.2) | (64, 64, 64) |
| 3 | Conv2D + BatchNomalization + LeakyReLU (alpha=0.2) | (32, 32, 128) |
| 4 | Conv2D + BatchNomalization + LeakyReLU (alpha=0.2) | (16, 16, 256) |
| 5 | Conv2D + BatchNomalization + LeakyReLU (alpha=0.2) | (8, 8, 512) |
| 6 | Conv2D | (8, 8, 1) |

*Layer structure uses the abbreviate terminology in TensorFlow 2.5.

# Appendix C

# Linear elastic fracture mechanics (LEFM) estimation of dynamic fracture toughness

From the linear elastic fracture mechanics (LEFM) theory, the Mode I crack-tip stress intensity factor as a function of time under a step loading of normal tensile wave can be expressed as (Freund, 1998),

$$K_I(t) = n(v)\sigma_0 C_l^{1/2} t^{1/2} \qquad (C.1)$$

where $n(v) = \frac{2}{1-v}\left(\frac{1-2v}{\pi}\right)^{\frac{1}{2}}$ with Poisson's ratio $v = 0.4$ is the blunt factor, $\sigma_0 = \frac{1}{2}\rho C_l V_0$ is the incident normal stress with density $\rho = 1070 \ kg/m^3$ and impact velocity $V_0 = 204 \ m/s$. $C_l = \left(\frac{E_G}{\rho(1-v^2)}\right)^{1/2} = 2100 \ m/s$ is the longitudinal wave speed. At a critical time $\tau$, the crack is assumed to initiate, the critical Mode I stress intensity factor $K_{IC}$ can be expressed as,

$$K_{IC} = n(v)\sigma_0 C_l^{1/2} \tau^{1/2} \qquad (C.2)$$

Therefore, the toughness $G_c$ under plane-strain condition is expressed as,

$$G_c = \frac{(1-v^2)K_{IC}^2}{E_G} \qquad (C.3)$$

where $E_G = 3.96 \ GPa$ is Young's modulus of the polyurea. From FEM, the incubation time of crack-growth initiation is $\tau = 0.8 \ \mu s$. The crack-growth initiation time is defined as the time between two events when the tensile wave arrives at the crack surface and when the separation of two crack-tip nodes reaches $\delta_c$. Therefore, the linear elastic fracture mechanics (LEFM) estimation of the toughness provides $G_c^{LEFM} = 13.4 \ kJ/m^2$.

Fernández-Godino, M.G., Park, C., Kim, N.-H., Haftka, R.T., 2016. Review of multi-fidelity models. arXiv preprint arXiv:1609.07196.

Freund, L., 1972. Crack propagation in an elastic solid subjected to general loading—I. Constant rate of extension. J Mech Phys Solids 20, 129-140.

Freund, L., 1973. Crack propagation in an elastic solid subjected to general loading—III. Stress wave loading. J Mech Phys Solids 21, 47-61.

Freund, L.B., 1998. Dynamic fracture mechanics. Cambridge university press.

Gao, H., Rice, J.R., 1989. A first-order perturbation analysis of crack trapping by arrays of obstacles. J Appl Mech 56(4), 828–836.

Goodfellow, I.J., Pouget-Abadie, J., Mirza, M., Xu, B., Warde-Farley, D., Ozair, S., Courville, A., Bengio, Y., 2014. Generative Adversarial Networks, p. arXiv:1406.2661.

Grujicic, M., Snipes, J.S., Galgalikar, R., Ramaswami, S., 2014. Material-Model-Based Determination of the Shock-Hugoniot Relations in Nanosegregated Polyurea. J Mater Eng Perform 23, 357-371.

Grujicic, M., Yavari, R., Snipes, J.S., Ramaswami, S., Jiao, T., Clifton, R.J., 2015. Experimental and Computational Study of the Shearing Resistance of Polyurea at High Pressures and High Strain Rates. J Mater Eng Perform 24, 778-798.

Grunschel, S.E., 2009. Pressure-Shear Plate Impact Experiments on High-Purity Aluminum at Temperatures Approaching Melt. Brown University, Ph.D. thesis.

Hong, S., Chew, H.B., Kim, K.-S., 2009. Cohesive-zone laws for void growth - I. Experimental field projection of crack-tip crazing in glassy polymers. J Mech Phys Solids 57, 1357-1373.

Hong, S.S., Kim, K.-S., 2003. Extraction of cohesive-zone laws from elastic far-fields of a cohesive crack tip: a field projection method. J Mech Phys Solids 51, 1267-1286.
48